\documentclass[useAMS,usenatbib]{mnras}
\usepackage[normalem]{ulem}

\usepackage{tabularx,ragged2e,booktabs}
\usepackage{natbib}
\usepackage{graphicx}
\usepackage{amsmath,amsfonts,amssymb}
\usepackage{multirow}
\usepackage{hyperref}
\usepackage{color}
\usepackage{cases}
\usepackage[figure]{hypcap}
\usepackage{bm}

\hypersetup{colorlinks, citecolor=blue, pdfduplex=DuplexFlipLongEdge}
\hypersetup{pdftitle=heating paper, pdfauthor=Andrew Chael, pdfsubject=Astrophysics}
\def\sgraa{{Sgr~A$^{\ast} $}}
\def\adi{\Gamma_\mathrm{gas}}

\def\bcdot{\boldsymbol{\cdot}}

\title[Electron heating in Sgr A*]{The role of electron heating  physics in images and variability of the Galactic Centre black hole Sagittarius A*}
\author[A. Chael et al.]
{Andrew Chael$^{1}$\thanks{\hbox{E-mail: achael@cfa.harvard.edu}},
Michael Rowan$^{1}$,
Ramesh Narayan$^{1}$,
Michael Johnson$^{1}$
\newauthor
and Lorenzo Sironi$^2$
\\
$^{1}$Harvard-Smithsonian Center for Astrophysics, 60 Garden Street, Cambridge, MA 
02138, USA\\
$^{2}$Department of Astronomy, Columbia University, 550 W 120th St, New York, NY 10027, USA\\
}

\begin{document}
\maketitle

\begin{abstract}
The accretion flow around the Galactic Centre black hole Sagittarius A* (\sgraa) is expected to have an electron temperature that is distinct from the ion temperature, due to weak Coulomb coupling in the low-density plasma. We present four two-temperature general relativistic radiative magnetohydrodynamic (GRRMHD) simulations of \sgraa\ performed with the code \texttt{KORAL}. These simulations use different electron heating prescriptions, motivated by different models of the underlying plasma microphysics. We compare the Landau-damped turbulent cascade model used in previous work with a new prescription we introduce based on the results of particle-in-cell simulations of magnetic reconnection. With the turbulent heating model, electrons are preferentially heated in the polar outflow, whereas with the reconnection model electrons are heated by nearly the same fraction everywhere in the accretion flow. The spectra of the two models are similar around the submillimetre synchrotron peak, but the models heated by magnetic reconnection produce variability more consistent with the level observed from \sgraa. All models produce 230~GHz images with distinct black hole shadows which are consistent with the image size measured by the Event Horizon Telescope, but only the turbulent heating produces an anisotropic `disc-jet' structure where the image is dominated by a polar outflow or jet at frequencies below the synchrotron peak. None of our models can reproduce  the observed radio spectral slope, the large near-infrared and X-ray flares, or the near-infrared spectral index, all of which suggest non-thermal electrons are needed to fully explain the emission from \sgraa. 
\end{abstract}


\section{Introduction}
\label{sec::intro}

Observations of Sagittarius A* (\sgraa), the $\sim4\times10^6 M_\odot$ black hole at the centre of the Milky Way,
offer a unique window on accretion physics, collisionless plasma processes, and gravity itself. In contrast to distant luminous active galactic nuclei,
the accretion flow surrounding \sgraa\
has an extremely low luminosity $\sim10^{-9}$ of Eddington \citep{Falcke98, Genzel03, Baganoff03}
and a low mass accretion rate $\sim10^{-7}$ of Eddington \citep{Agol2000, Bower2003, Marrone_2007}. 

The low luminosity and spectrum of \sgraa\ and similar systems have been successfully explained within the context of analytic
advection-dominated accretion flow models \citep[ADAFs:][]{Ichimaru77, Rees1982, Narayan_Yi, Narayan_Yi_Nat, Narayan98, Maha98, Blandford99}.
In these systems, most of the gravitational potential energy liberated 
by the infall of gas is lost by advection across the black hole event horizon or in an outflow \citep[see the review by][]{Yuan14}. Consequently, the radiative luminosity of these systems is very low. Forthcoming spatially resolved observations at 230~GHz by the Event Horizon Telescope (EHT) \citep{Doeleman08} will produce images of the submillimetre emission from the \sgraa\ accretion flow within a few Schwarzchild radii of the black hole. 
Observations with the GRAVITY interferometer \citep{Gillessen10} will track the near-infrared centroid during flares with similar precision. 

While analytic models can successfully explain elements of the spectrum of \sgraa, they are typically one-dimensional and rely on artificial 
viscosity for angular momentum transport. They are therefore not suitable for detailed comparisons with \sgraa's spectrum and rapid variability. 
These features are most easily explored via simulations. 
Fortunately, ADAF models are optically thin and geometrically thick, making them particularly 
tractable for grid-based magnetohydrodynamic simulations. 
Multiple general relativistic magnetohydrodynamic (GRMHD) and radiative magnetohydrodynamic (GRRMHD) codes have been developed to simulate the plasma flow around black holes
\citep[e.g.][]{Komissarov99,DeVilliers03b,Gammie03,Tchekhovskoy07,KORAL13,KORAL14,McKinney14,Ryan15}. 
These codes have been successfully applied to simulations of low accretion rate systems like Sgr A*
\citep[e.g.][]{Hawley00,DeVilliers03b,Gammie03, Tchekhovskoy10, McKinney2012, Narayan2012} and have been successfully used to demonstrate that viscosity in these collisionless systems
can arise by turbulence generated by the magnetorotational instability \citep[MRI:][]{BalHawley,DeVilliers03b,Narayan2012}.

In hot ionized systems like Sgr A*, Coulomb coupling
between electrons and ions is inefficient. Electrons and ions will therefore
have distinct temperatures. Knowledge of the electron temperature is necessary to produce images and spectra from simulations for comparison to observations. Often, the electron-to-gas
temperature ratio is set manually after the simulation. Usually $T_\mathrm{e}/T_\mathrm{gas}$ is fixed to a constant value,
\citep[e.g.][]{Moscibrodzka_09, Dexter10}, but some authors have divided the simulation into jet and disc
regions with different temperature ratios \citep[e.g][]{Moscibrodzka_13,Moscibrodzka_14,Chan_15a, Chan_15b},
or computed temperature ratios based on fluid properties in the midplane \citep[e.g.][]{Shch_12}.
 
\citet{Ressler15} and \citet{KORAL16} have
developed codes which self-consistently evolve a thermal electron population alongside the other fluid variables. 
In the method of \citet{Ressler15}, a single fluid accretion flow is simulated with a standard GRMHD code. Dissipation in the simulation is then identified and applied as heating in the evolution the electron entropy in a post-processing step. In \citep{Ressler17}, this method successfully produced a model of \sgraa in line with observations of the quiescent spectrum and variability properties. 
\citet{KORAL16} built a more general framework in which ions and electrons are evolved simultaneously along with the total gas and radiation in a GRRMHD simulation, including
radiative and Coulomb couplings (although the latter is not important for \sgraa). In this method, the electron temperature is obtained `on the fly' during the simulation and not computed during post-processing. Recently, \citet{Ryan17} have adapted the method of \citet{Ressler15} for self-consistent evolution of a thermal electron population in a GRRMHD simulation with frequency-dependent radiative transport  \citep{Ryan15}.

While the physics of electron cooling (due to synchrotron, free-free, and inverse Compton emission) is well understood, the physics of electron heating
in collisionless accretion flows is largely unconstrained. 
As the gas falls in toward the black hole, it is heated by both adiabatic compression and viscous dissipation. In a simulation,
the total amount of viscous dissipation generated in each spatial cell at each timestep can be computed numerically by comparing the energy generated in the total gas evolution
to the adiabatically evolved entropies of the different component fluids. However, the \emph{fraction} 
of the viscous heat that goes into the electrons is determined by microphysical processes beyond the reach of the GRMHD simulation. The best one can do is to model this microphysical heating via a sub-grid prescription.

Past works \citep{Ressler15, KORAL16, Ressler17, Ryan17} have used a heating prescription based on the Landau-damped cascade of weakly collisional MHD turbulence \citep{Howes10}. This model was originally developed
for solar wind observations, and it is unclear if it is applicable in the environment around \sgraa, 
where electrons are expected to be nearly or entirely relativistic.

It has recently been suggested that magnetic reconnection
may be a critical element of MHD turbulence \citep{Carbone1990,Boldyrev2017,Loureiro2017, Mallet2017, Comisso2018}. In the
dynamic alignment picture \citep{Boldyrev2006}, turbulent
eddies become progressively more sheet-like, and more susceptible to the tearing
mode instability which drives reconnection as their characteristic
scale gets smaller. One would then expect that energy dissipation in MHD turbulence is ultimately mediated by reconnection at small scales. Motivated by this fact, \citet{Rowan17} explored magnetic reconnection
as an alternative physical origin of electron heating for a range of plasma parameters closer to those expected in the gas around \sgraa (see also \citet{Werner2018} for a similar study). 
They performed a suite of particle-in-cell simulations of anti-parallel reconnection (i.e., in the absence of a guide field perpendicular to the alternating fields), and they measured the relative 
amounts of the resulting electron and ion heating as a function of the local plasma conditions.

In this paper we present the results of four 3D two-temperature simulations of \sgraa\ using the code \texttt{KORAL} \citep{KORAL13,KORAL14,KORAL16}. In these simulations, we compare the Landau-damped turbulent cascade heating prescription from \citet{Howes10} with a new prescription for magnetic reconnection heating that is fit to particle-in-cell simulation data from \citet{Rowan17}. We perform a simulation for each heating prescription at both low ($a=0$) and high ($a=0.9375$) black hole spins.  

In Section~\ref{sec::phys}, we review the method of \citet{KORAL16} for evolving
thermal electron and ion entropies alongside the other fluid variables.  
We also introduce the two heating prescriptions considered in this paper and describe their dependence on 
plasma parameters. In Section~\ref{sec::num}, we describe the setup of our numerical simulations and 
our method for computing spectra and images from the simulation outputs. In Section~\ref{sec::results} we present the results of 
our simulations. We discuss the dynamics, thermodynamics, and magnetization of the accreting gas, and we describe the predicted spectra, lightcurves, and 230~GHz images 
such as may soon be observed by the Event Horizon Telescope \citep{Doeleman08}.
In Section~\ref{sec::discussion} we compare our results with those of \citet{Ressler17} and discuss their implications for future models of \sgraa. We conclude in Section~\ref{sec::summary}.

We note that, in \sgraa, electron-electron collisions may not be sufficient to entirely relax the electron distribution function to a thermal Maxwellian \citep{Maha97}. Shocks and magnetic reconnection can
accelerate a fraction of the electrons into relativistic non-thermal distributions 
which persist alongside the  lower-energy thermal distribution. In \citet{Chael17}, we introduced an extension of the two species thermal approach (\citet{KORAL16} and this work) to evolve arbitrary electron distributions in accretion simulations. We found that localized non-thermal particle injection is likely necessary to produce the extreme infrared and X-ray flares observed in Sgr A* \citep{Ball_16}. The present work confirms this conclusion. A forthcoming work will explore localized non-thermal particle acceleration models in the context of self-consistent accretion simulations as candidates for the origin of these flares.

\section{Electron and Ion Thermodynamics}
\label{sec::phys}
\subsection{Evolution Equations}
In this section, we review the method used in the GRRMHD code \texttt{KORAL} for evolving a two-temperature fluid \citep{KORAL16}. Standard GRMHD simulations \citep[e.g.][]{Gammie03} track a magnetic field four-vector $b^\mu$ coupled to a single ionized
perfect fluid (assumed for the remainder of this work to be pure Hydrogen), which is characterized by a gas density 
$\rho$, internal energy density $u$, and four-velocity $u^\mu$. 
The pressure $p$ is determined by the internal energy $u$ and the adiabatic index $\adi$, $p=(\adi-1)u$.
The adiabatic index is typically fixed at the non-relativistic monatomic value $\adi = 5/3$. The perfect fluid
is evolved using the conservation of rest mass and stress-energy, combined with the ideal MHD induction equation.

Radiative general relativistic MHD simulations (GRRMHD) add an additional massless fluid to represent the frequency-integrated
radiation field. Under the M1 closure scheme used in \citet{KORAL13,KORAL14,McKinney14}, the radiation field is 
completely determined by a rest frame radiation energy density $\bar{E}$ and a
timelike four velocity $u^\mu_\mathrm{r} \neq u^\mu$ of the frame where the radiation is isotropic. The coupling 
between the radiation and gas is determined by frequency-averaged opacities $\kappa$ due to synchrotron and
bremsstrahlung emission, and inverse Compton scattering. Furthermore, by introducing the photon number density $\bar{n}_\mathrm{r}$ 
as an additional simulation variable, we can obtain the radiation temperature $T_\mathrm{r}$ (roughly corresponding
to the peak frequency) under the assumption that the radiation spectrum is a grey body \citep{Compt15}.

\citet{KORAL16} introduced \emph{two-temperature} GRRMHD to \texttt{KORAL}. In this method, we assume both electrons and ions share the same 
fluid velocity $u^\mu$, and use charge neutrality to set the number densities $n_\mathrm{e} = n_\mathrm{i} = n = \rho/m_\mathrm{i}$, where $m_\mathrm{i}$ is the proton mass.  
We then evolve two new fluid variables: the entropy per particle of electrons ($s_\mathrm{e}$) and ions ($s_\mathrm{i}$). 
The species rest-frame temperatures $T_\mathrm{e}$ and $T_\mathrm{i}$ 
are functions of the entropy per particle $s_\mathrm{e}, s_\mathrm{i}$ and density (see Appendix~\ref{appendix::eqs}). 
Because electrons in hot accretion flows can be nearly or entirely relativistic, we include for each species an adiabatic index 
$\Gamma_{\rm e,i}(\Theta_{\rm e,i})$ that depends on the dimensionless temperature $\Theta_{\rm e,i} = k_{\rm B} T_{\rm e,i}/m_{\rm e,i}c^2$, where $m_\mathrm{e},m_\mathrm{i}$ are the 
electron and proton mass, respectively. For non-relativistic particles ($\Theta_{\rm e,i} \ll 1$),
$\Gamma_{\rm e,i}=5/3$, while for relativistic particles ($\Theta_{\rm e,i} \gg 1$), $\Gamma_{\rm e,i}=4/3$. The species energy densities are then given by 
\begin{equation}
u_{\rm e,i} = \frac{n k_{\rm B} T_{\rm e,i}}{\Gamma(\Theta_{\rm e,i})-1}.
\end{equation}
At the end of every timestep, the two species energy densities must add to the separately evolved total gas energy density $u$. The adiabatic index of the
combined fluid is a combination of the two species adiabatic indices (see Appendix~\ref{appendix::eqs}).

The first law of thermodynamics determines the evolution of the species entropies. In covariant form, this is
\begin{align}
\label{eq::ent_ev}
T_\mathrm{e}\left(n s_\mathrm{e} u^\mu\right)_{;\mu} &= \delta_\mathrm{e} q^{\rm v} + 
q^{\rm C} - \hat{G}^0, \\
\label{eq::ent_ev2}
T_\mathrm{i}\left(n s_\mathrm{i} u^\mu\right)_{;\mu} &= (1-\delta_\mathrm{e}) q^{\rm v} - q^{\rm C},
\end{align}
where $q^\mathrm{v}$ is the total dissipative heating rate, $\delta_\mathrm{e}$ is the fraction of the dissipative heating that goes into electrons, $q^\mathrm{C}$ is the Coulomb coupling energy exchange rate, and $\hat{G}^0$ is the radiative cooling rate.  

In the absence of dissipative heating and cooling (the right sides of equations~\ref{eq::ent_ev} and \ref{eq::ent_ev2} set to zero), the species entropies are conserved and the particles heat and cool only due to adiabatic compression and expansion. The first source term 
on the right hand side of both equations~\eqref{eq::ent_ev} and \eqref{eq::ent_ev2} is
the dissipative heating. Dissipative heating arises in accretion 
systems at scales far smaller than the grid-scale by physical processes which may include
turbulent damping, magnetic reconnection, and shock heating. In our simulations, we cannot resolve these processes but we can
identify the total dissipative heating rate $q^{\rm v}$ numerically at the grid scale. 

To numerically identify the total dissipative heating, after evolving the fluid through a proper time step $\Delta \tau$, we evolve the thermal entropies adiabatically 
(solving equations~\ref{eq::ent_ev}--\ref{eq::ent_ev2} with the right hand sides set to zero). We then compute the adiabatically
evolved energy densities $u_{\rm i,\,adiab}$ and $u_{\rm e,\,adiab}$ (using the relations in Appendix~\ref{appendix::eqs}), 
and then compute the total dissipation by comparing their sum to the total gas energy:
\begin{equation}
 \label{eq::vischeat}
  q^{\rm v} = \frac{1}{\Delta \tau}\left[u - u_{\rm i\,adiab} - u_{e\,adiab}\right].
\end{equation}

While the total viscous heating rate can be computed numerically, the fraction of the heating that goes into the electrons, $\delta_\mathrm{e}$,
must be determined from sub-grid physics as a function of the local plasma parameters. In section~\ref{sec::heating}, we discuss the two  
heating prescriptions (functions for $\delta_\mathrm{e}$) that we explore in this paper.

The second source term on the right hand sides of equations~\eqref{eq::ent_ev} and \eqref{eq::ent_ev2} is the (weak) Coulomb coupling $q^{\rm C}$ \citep{Stepney83}.
Finally, $\hat{G}^0$ is the net electron energy lost to absorption and emission of radiation. In this work, we consider synchrotron, free-free, and inverse Compton contributions to $\hat{G}^0$. 

\texttt{KORAL} solves equations~\eqref{eq::ent_ev} and \eqref{eq::ent_ev2} for the electron thermodynamics in parallel with conservation equations for the total matter and radiation fluids and the induction equation for the magnetic field using a split explicit-implicit scheme. The advection of quantities across cells is handled explicitly by reconstructing the appropriate fluxes at the cell walls. The source terms in the evolution equations which represent the coupling between matter and radiation or between particle species are applied implicitly at each cell center \citep{KORAL13,KORAL14,KORAL16}.

\subsection{Electron Heating Prescriptions}
\label{sec::heating}
\begin{figure}
\includegraphics*[width=0.45\textwidth]{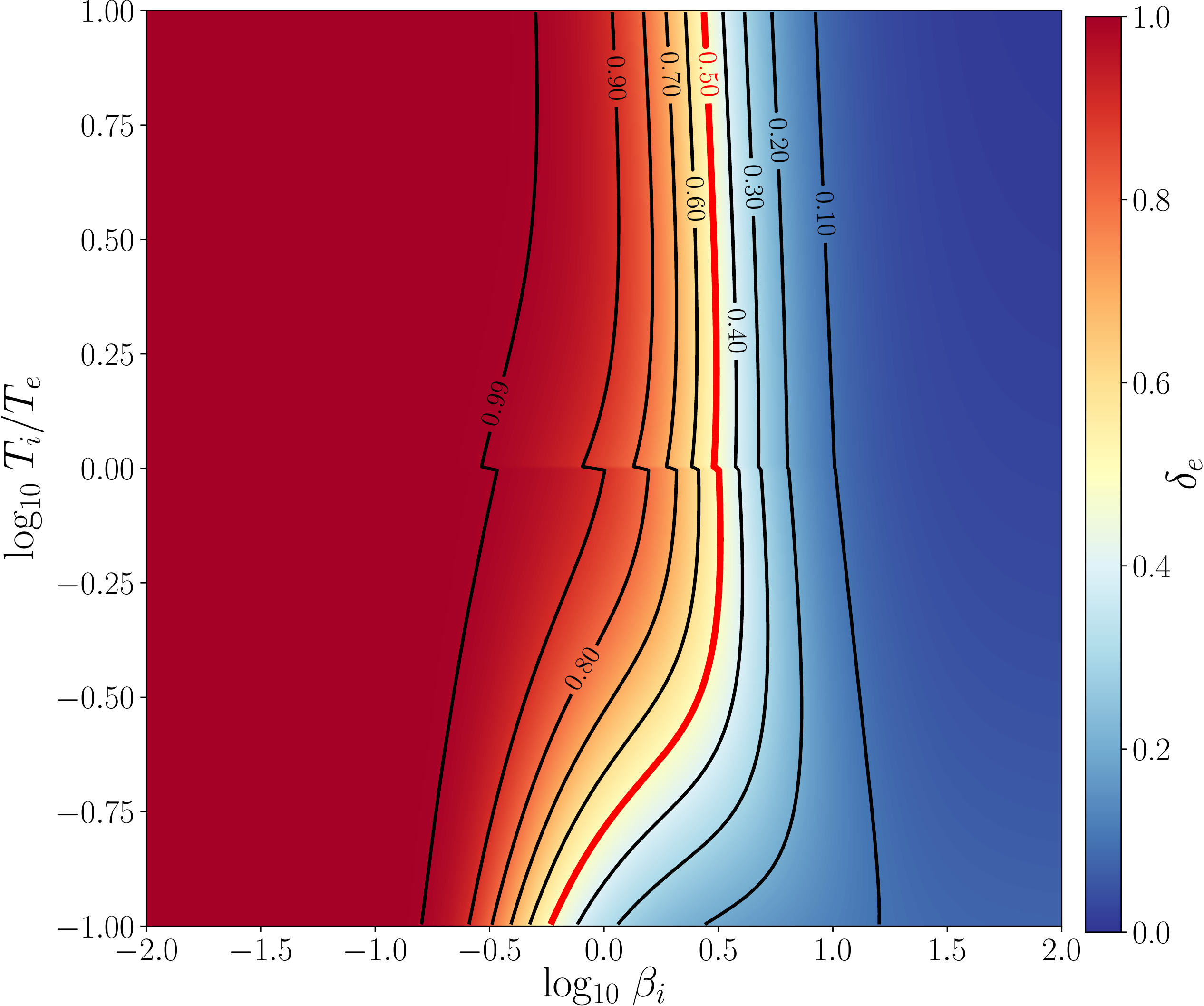}
\caption{
Turbulent cascade heating prescription \citep{Howes10}. The electron heating fraction $\delta_\mathrm{e}$ (i.e., the fraction of dissipated energy converted to irreversible electron heating) is plotted as a function of $\beta_\mathrm{i}$ and $T_\mathrm{i}/T_\mathrm{e}$. At all values of $T_\mathrm{i}/T_\mathrm{e}$, this prescription transitions rapidly from putting most of the dissipated energy into electrons at low $\beta_\mathrm{i}$ to putting most of the dissipated energy into ions at high $\beta_\mathrm{i}$. The red contour denotes $\delta_\mathrm{e} = 0.5$.
}
\label{fig::howesmodel}
\end{figure}

In this work, we examine simulations that use two different sub-grid prescriptions for $\delta_\mathrm{e}$, the fraction 
of the local dissipative energy generated by the simulation that heats the electrons. The value of $\delta_\mathrm{e}$ is 
determined by plasma physics processes that occur below the grid scale. In our simulations, we explore two different 
physical prescriptions which determine $\delta_\mathrm{e}$ based on the local plasma physics parameters $\beta_\mathrm{i}$, $\sigma_\mathrm{i}$,
and the temperature ratio $T_\mathrm{i}/T_\mathrm{e}$. 

The parameter $\beta_\mathrm{i}$  is the ratio of the thermal ion pressure to the magnetic pressure:
\begin{equation}
 \label{eq::beta}
 \beta_\mathrm{i} = \frac{8\pi \, n_\mathrm{i} k_{\rm B} T_\mathrm{i}}{|B|^2}.
\end{equation}
In our simulations $\beta_\mathrm{i} \sim10$ in the midplane, but above and below the midplane $\beta_\mathrm{i}$ drops to $\beta_\mathrm{i} \sim1$, and
in the jet region close to the axis $\beta_\mathrm{i} < 1$. 

The magnetization $\sigma_\mathrm{i}$ compares the magnetic energy to the rest-mass energy of the fluid:
\begin{equation}
 \label{eq::sigmai}
 \sigma_\mathrm{i} = \frac{|B|^2}{4\pi \, n_\mathrm{i} m_\mathrm{i} c^2}.
\end{equation}
In our simulations $\sigma_\mathrm{i} < 1$ everywhere except in the innermost jet region, but it
is still relatively high compared to more familiar environments
such as the non-relativistic solar wind ($\sigma_\mathrm{i} \ll 1$).
In the accretion disc close to the black hole ($r < 25\, r_{\rm g}$), $\sigma_\mathrm{i}$ is generally in the range $1 > \sigma_\mathrm{i} > 10^{-3}$. 

All previous two-temperature GRMHD studies of \sgraa\ have used the turbulent heating prescription from \citet{Howes10}. This prescription is based on
calculations of the Landau damping of a MHD turbulent cascade in a weakly collisional plasma \citep{Howes2008a} with $\sigma_\mathrm{i} \ll 1$, and Kolmogorov constants were determined by fitting to kinetic simulations \citep{Howes2008b}. The Howes prescription matches well to solar wind measurements of the electron-to-ion heating rates \citep{Howes11}. The full Howes fitting function is
\begin{align}
 \label{eq::Howes}
 \delta_\mathrm{e} &= \frac{1}{1+f_\mathrm{e}}, \\
 f_\mathrm{e} &= c_1 \frac{c_2^2 + \beta_{\rm i}^{2-0.2\log_{10}(T_\mathrm{i}/T_\mathrm{e})}}{c_3^2 + \beta_{\rm i}^{2-0.2\log_{10}(T_\mathrm{i}/T_\mathrm{e})}}\sqrt{\frac{m_\mathrm{i}T_\mathrm{i}}{m_\mathrm{e}T_\mathrm{e}}}\mathrm{e}^{-1/\beta_\mathrm{i}},
\end{align}
where $c_1 = 0.92$ and $c_2 = 1.6\,T_\mathrm{e}/T_\mathrm{i},\; c_3 = 18 + 5\log_{10}(T_\mathrm{i}/T_\mathrm{e})$ if $T_\mathrm{i}>T_\mathrm{e}$, and  $c_2 = 1.2\,T_\mathrm{e}/T_\mathrm{i},\; c_3 = 18$ if $T_\mathrm{i} < T_\mathrm{e}$.

The Howes turbulent cascade prescription is a weak function of the temperature ratio $T_\mathrm{e}/T_\mathrm{i}$ but a strong function of $\beta_\mathrm{i}$ (see Fig.~\ref{fig::howesmodel}).
It gives most of the turbulent heating to electrons at low $\beta_\mathrm{i}$, and conversely gives most of the heating to ions at high $\beta_\mathrm{i}$. This is a general result predicted for damped MHD turbulence \citep{Quataert99}. 
Since $\delta_\mathrm{e} \approx 1$ in regions of low $\beta_\mathrm{i}$, when the Howes turbulent cascade prescription is applied to accretion simulations, we expect the resulting electron temperature to be higher in the polar/jet region (where $\beta_\mathrm{i}<1$) when compared to the equatorial plane (where $\beta_\mathrm{i}>1$).

\begin{figure}
\includegraphics*[width=0.45\textwidth]{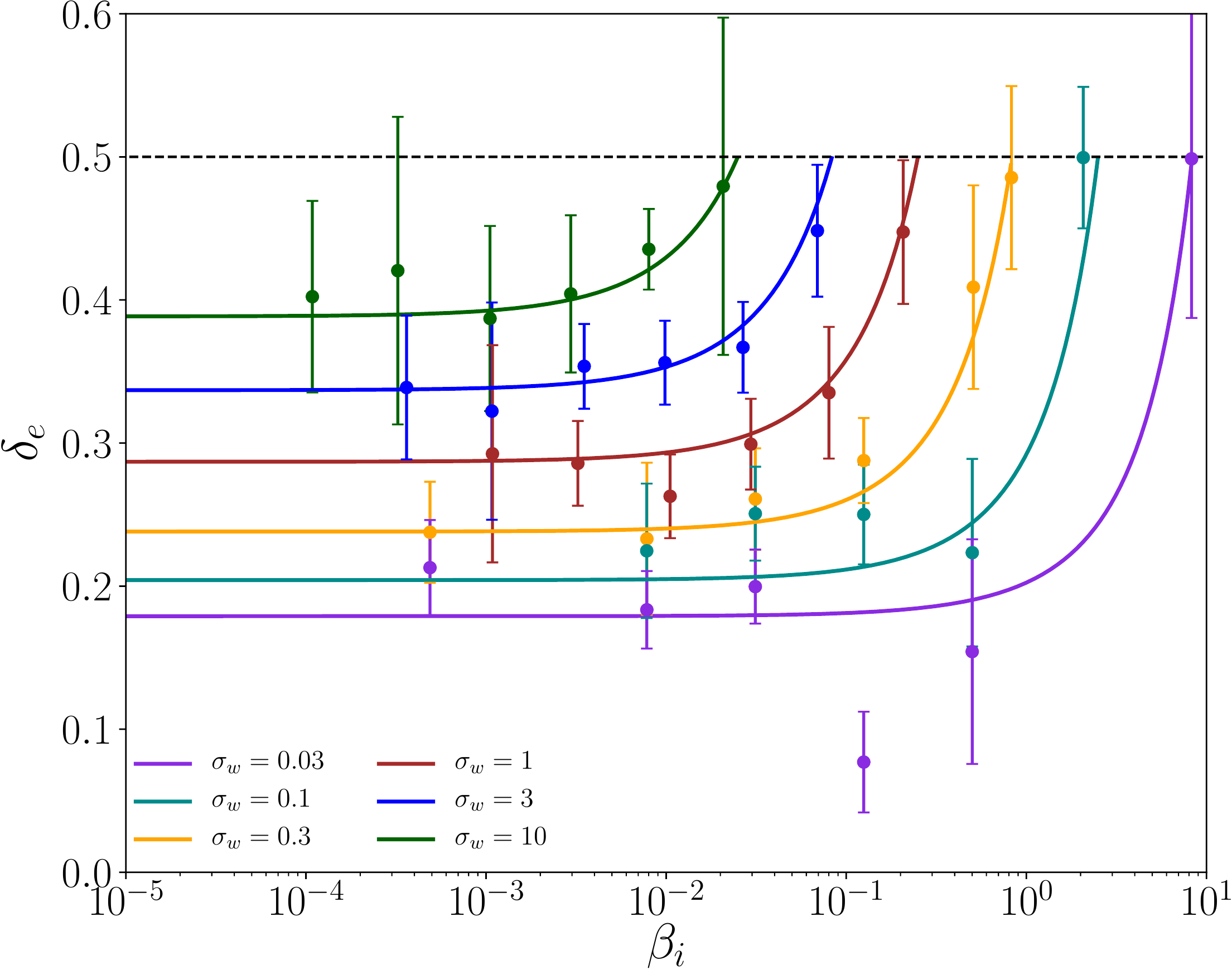}
\caption{
Reconnection heating prescription (equation~\ref{eq::Rowan}) fit to PIC simulation data from \citet{Rowan17}. The irreversible heating fraction $\delta_{e}$ is plotted against $\beta_{i}$ for various values of magnetization $\sigma_{w},$ indicated by the color.  Points show the results of simulations and lines show a fit  to these data. Error bars indicate, roughly, $\pm1.0 \, \sigma$ confidence intervals on the heating fractions extracted from each PIC simulation. For a given $\sigma_w$, the maximum allowed $\beta_\mathrm{i}$, where the ions and electrons both are ultra-relativistic, is $\beta_{\rm i,\text{max}} = 1/4\sigma_w$. The functional form we use constrains the heating fraction $\delta_\mathrm{e} \rightarrow 0.5$ as $\beta_\mathrm{i} \rightarrow \beta_{\rm i,\text{max}}$.
}
\label{fig::rowanmodel_data}
\end{figure}

\begin{figure*}
\centering
\includegraphics*[width=0.9\textwidth]{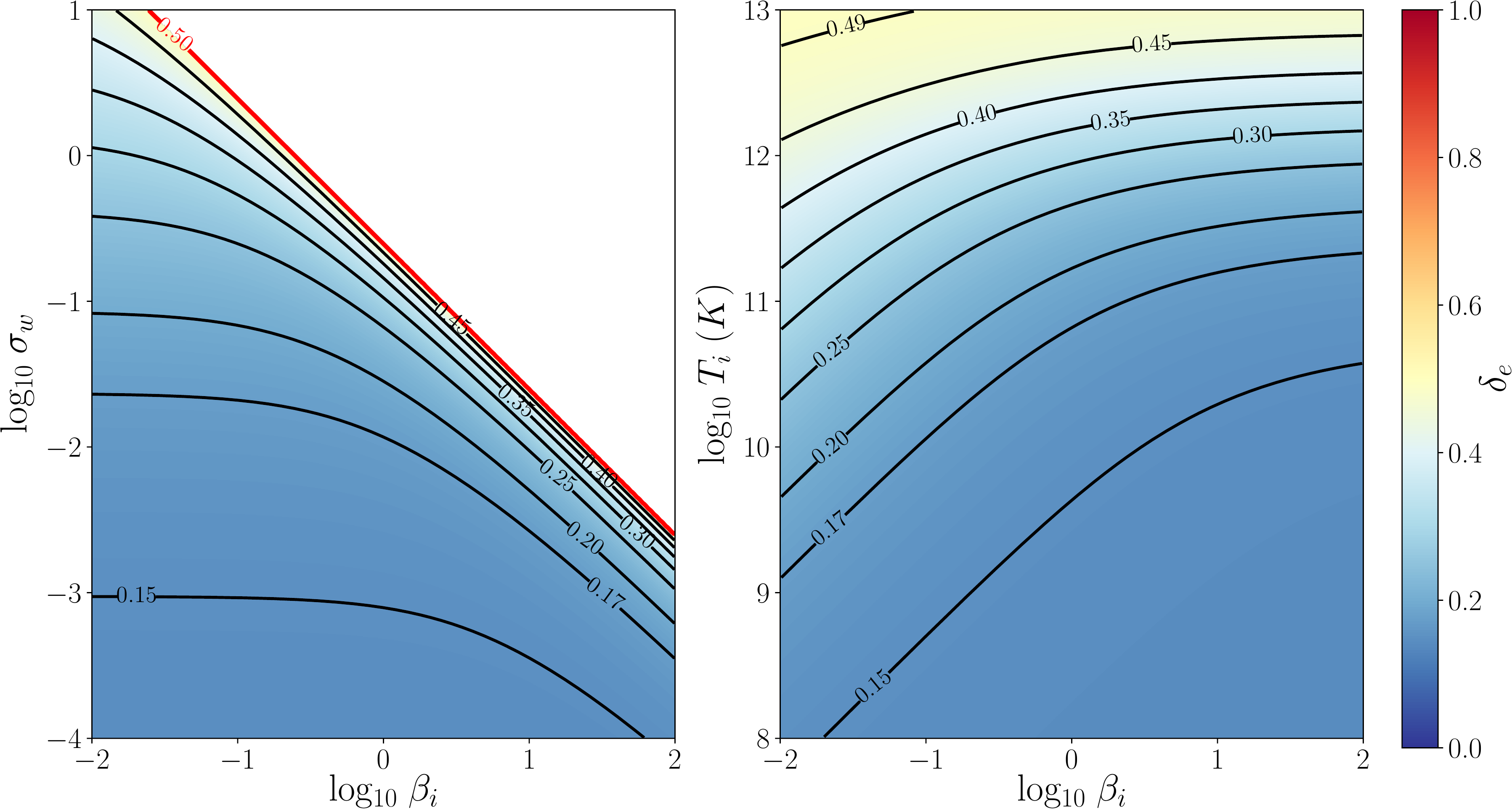}
\caption{Magnetic reconnection heating prescription (equation~\ref{eq::Rowan}). (Left) $\delta_\mathrm{e}$ plotted as a function of the independent variables $\beta_\mathrm{i}$ and $\sigma_w$. 
The red contour denotes $\delta_\mathrm{e} = 0.5$,
which occurs only at $\beta_{\rm i,\text{max}} = 1/4\sigma_w$.
(Right) $\delta_\mathrm{e}$ plotted as a function of $\beta_\mathrm{i}$ and $T_\mathrm{i}$ in Kelvin, assuming $T_\mathrm{i}=T_\mathrm{e}$.
Most of the dissipated energy goes into ions at low $\beta_\mathrm{i}$, 
and $\delta_\mathrm{e} \rightarrow 0.5$ at high temperatures. 
}
\label{fig::rowanmodel}
\end{figure*}

It has recently been suggested that the nature of MHD turbulence at small scales (where the dissipation happens) is modified by magnetic reconnection  \citep{Carbone1990, Boldyrev2017, Loureiro2017, Mallet2017,Comisso2018}. Turbulent eddies become sheet-like, and they naturally fragment into plasmoids/magnetic islands via  the tearing mode instability of reconnecting current sheets. One would then expect that energy dissipation in MHD turbulence is ultimately mediated by reconnection. For this reason, we compare the Howes prescription for particle heating from turbulent dissipation to one developed from measurements of electron heating in fully kinetic particle-in-cell (PIC) simulations of trans-relativistic reconnection presented in \citet{Rowan17}.  In these PIC simulations, the strength of the magnetic field is parametrized by the magnetization as defined with respect to the fluid enthalpy $w$:
\begin{align} \label{eq:sigw}
\sigma_{w} = \frac{|B|^{2}}{4\pi\,w} = \frac{|B|^2}{4 \pi \, (n_{\rm i} m_{\rm i} c^2 + \Gamma_{\rm i} u_{\rm i} + \Gamma_{\rm e} u_{\rm e})},
\end{align}
where the adiabatic indices $\Gamma_{\rm e}, \Gamma_{\rm i}$ are functions of dimensionless temperatures $\Theta_{\rm e}, \Theta_{\rm i}$ respectively (see Appendix~\ref{appendix::eqs}).

If we assume $T_\mathrm{e} = T_\mathrm{i}$ we can write $\sigma_w$ as
\begin{equation}
\label{eq::sigmaweq}
\left.\sigma_w\right|_{T_\mathrm{e}=T_\mathrm{i}} = \frac{2}{\beta_{\rm i}\left(\Theta_\mathrm{i}^{-1} + \frac{\Gamma_{\rm i}}{\Gamma_{\rm i} -1} + \frac{\Gamma_{\rm e}}{\Gamma_{\rm e} -1}\right)}.
\end{equation}
From equation~\eqref{eq::sigmaweq}, we can see that for a given $\sigma_w$, $\beta_\mathrm{i}$ has a maximum value $\beta_{\rm i,\text{max}} = 1/4\sigma_w$ which is achieved in the limit when both species are highly relativistic and the thermal energy dominates the rest mass energy ($\Theta_\mathrm{i} \gg 1, \; \Gamma_{\rm i,e}\rightarrow 4/3$).

The magnetization $\sigma_w$ represents the initial magnetic energy per electron-proton pair in units of initial particle enthalpy.  Through the reconnection of  magnetic fields, some of this initial magnetic energy can be transferred from the electromagnetic field to particles, as plasma passes from the pre-reconnection `upstream' to the post-reconnection`downstream' region. In the simulations of \citet{Rowan17}, plasma is initialized with a given temperature ratio $T_{\rm e}/T_{\rm i}$, magnetization $\sigma_{w},$ and plasma-beta $\beta_{\rm i}$ in the upstream region; the heating of electrons and protons is assessed by comparing the internal energy of particles in the upstream to those in the reconnection outflows.  

\citet{Rowan17} present results for both species for both the \emph{total} heating across the reconnection region and the \emph{irreversible} heating, which represents a genuine increase in the species entropy. In their treatment, the adiabatic heating $\Delta u_{\rm \,,adiab}$ is obtained by integrating the adiabatic first law of thermodynamics ($dU = -p dV$) across the reconnection region (see their eq. 20 for details). The irreversible heating is then obtained by subtracting off the adiabatic compressional heating from the total heating identified in their simulations. That is, if $u_{\rm e\,, down}$ and $u_{\rm i\,, up}$ are the electron energy densities measured downstream and upstream of reconnection, the irreversible heating electron heating is 
\begin{equation}
\label{eq::irrheat}
 \Delta u_{\rm e\,,irr} = (u_{\rm e\,, down} - u_{\rm e\,, up}) - \Delta u_{\rm e\,,adiab}.
\end{equation}
and the ratio of the \textit{irreversible} electron heating to the combined irreversible electron and ion heating is
\begin{equation}
 \label{eq::deltaeRowan}
 \delta_e = \frac{\Delta u_{\rm e\,,irr}}{\Delta u_{\rm e\,,irr} +  \Delta u_{\rm i\,,irr}}.
\end{equation}

We note that computing the irreversible, dissipative heating using Eq.~\ref{eq::irrheat} assumes that adiabatic compression and dissipation occur at distinct times and that compression occurs entirely before dissipation. In reality, both dissipation and compression occur simultaneously across the reconnection region, so the actual amount of dissipation generated is path-dependent. To investigate the dependence of the irreversible heating ratios $\delta_e$ calculated with the method of Eq.~\eqref{eq::deltaeRowan} used in \citet{Rowan17}, we performed a test in the opposite limit. Taking the temperatures and particle densities before and after reconnection from the simulations in \citet{Rowan17}, we computed how much dissipation would be needed for each species if all dissipation occurred \emph{before} compression. Presumably, the physical situation is bounded by these two extreme cases where dissipation occurs either entirely before or entirely after compression. Across all the PIC simulations considered, the change in $\delta_e$ was on the order of $\sim$20 per cent. The overall trend of $\delta_e$ with $\sigma_w$ and $\beta_i$ is unchanged, and the sharp qualitative difference between the numbers from \citet{Rowan17} simulations and the results of \citet{Howes10} persists.


We use the measured irreversible heating ratios $\delta_e$ from \citet{Rowan17} and derive a fitting function  with a similar functional form and the same asymptotic behavior as the fit to the total heating ratio presented in equation (34) of \citet{Rowan17}. Our reconnection heating prescription is thus
\begin{align} 
\label{eq::Rowan}
\delta_{\rm e} = \frac{1}{2} \exp \left[ \frac{-(1-\beta_{\rm i}/\beta_{\rm i, max})}{0.8+\sigma_{w}^{0.5}} \right],
\end{align}
where $\beta_{\rm i} < \beta_{\rm i, max} = 1/4\sigma_{w}.$  This fit has expected behavior in cases when the downstream scale separation between electrons and protons is of order unity. For example, when $\beta_{\rm i} \rightarrow \beta_{\rm i, max}$ or $\sigma_{w} \gg 1$.  When $\beta_{\rm i} \rightarrow \beta_{\rm i, max}, \delta_{\rm e} = 0.5$ for any value of $\sigma_{w}.$  Similarly, when $\sigma_{w} \gg 1,$ $\delta_{\rm e}=0.5,$ independent of $\beta_{\rm i}.$   

In Fig.~\ref{fig::rowanmodel_data}, we show the irreversible heating fractions $\delta_{e}$ as measured in PIC simulations from \citet{Rowan17}. The initial conditions of the upstream plasma in these simulations span the trans-relativistic regime of reconnection.  In all PIC simulation data presented, the upstream temperature ratio is fixed to unity $(T_{\rm e}/T_{\rm i} = 1)$ and the mass ratio is the physical $m_{\rm i}/m_{\rm e}=1836.$  For these PIC simulations, the ratio of ion thermal to magnetic pressure spans the range $10^{-4} < \beta_{\rm i} < 10$, and the magnetization ranges from $0.03<\sigma_w<10$.  Each curve in Fig.~\ref{fig::rowanmodel_data}, corresponding to a particular value of $\sigma_{w},$ is plotted as a function of $\beta_{\rm i}$ up to its maximum possible value $\beta_{\rm i, max} = 1/4 \sigma_{w}$.  The dashed black line at $\delta_{\rm e}=0.5$ indicates the limit for which electrons and ions have comparable heating efficiencies, $\delta_{\rm e}\rightarrow 0.5$. Although the data in Fig.~\ref{fig::rowanmodel_data} refer to simulations initialized with $T_\mathrm{e}/T_\mathrm{i}=1$, we have found that the electron and proton heating efficiencies are nearly independent of the upstream temperature ratio, in the regime $0.1 \leq T_\mathrm{e}/T_\mathrm{i}\leq 1$ we explored.

Fig.~\ref{fig::rowanmodel} shows that equation~\eqref{eq::Rowan} has qualitatively distinct behavior from the Howes turbulent cascade heating prescription, equation~\eqref{eq::Howes}. At fixed $\sigma_w$, taking $\beta_\mathrm{i} \rightarrow \beta_{\rm i,\, \text{max}}$ leads to $\delta_\mathrm{e} \rightarrow 0.5$, while taking high magnetizations $\sigma_w \gg 1$ also leads to $\delta_\mathrm{e} \rightarrow 0.5$. Plotting equation~\eqref{eq::Rowan} as a function of $T_\mathrm{i}$ and $\beta_\mathrm{i}$, making the assumption that $T_\mathrm{i}=T_\mathrm{e}$, we see that in the regime of interest for our simulations ($T_\mathrm{e} \sim 10^{10}\text{--}10^{12}$ K), $\delta_\mathrm{e}$ is relatively low, $\delta_\mathrm{e}\approx0.2\text{--}0.3$.  In contrast to the turbulent heating model, $\delta_\mathrm{e}$ never exceeds 0.5, indicating
that we should never expect $T_\mathrm{e}>T_\mathrm{i}$ in accretion simulations using this model. For values $\beta_{\rm i} \ll \beta_{\rm i, max},$ the irreversible heating fraction approaches a constant value that depends only on the magnetization;  this asymptotic value of $\delta_{\rm e}$ at $\beta_{\rm i} \ll \beta_{\rm i, max}$ decreases with the magnetization, as shown in Fig.~\ref{fig::rowanmodel}.  In the limit of non-relativistic reconnection ($\sigma_{w} \ll 0.1$), our fitting function yields $\delta_{\rm e} \rightarrow 0.14,$ which is consistent with the expectation that the heating fraction in the nonrelativistic reconnection limit is independent of magnetization. This is in rough agreement with recent laboratory experiments \citep{Eastwood13, Yamada14} and spacecraft observations \citep{Phan13, Phan14}.

In all PIC simulations used here, the magnetization $\sigma_{w} \ge 0.03,$ while the accretion simulations presented in this work have $\sigma_{w}\lesssim10^{-3}$ in the midplane of the accretion disc at radii larger than $r \gtrsim 25 \, r_{g}.$  If the behavior of $\delta_{e}$ from reconnection changes in the low $\sigma_{w}$ regime, this will have a major effect on the results of our global two-temperature Sgr A$^{*}$ simulations.  Future PIC studies are needed to investigate electron heating from reconnection at low magnetization, $\sigma_{w} < 0.03.$

In addition, our PIC simulations focused on the case of anti-parallel reconnection.  This may explain in part the discrepancy between the heating efficiencies quoted in \citet{Rowan17} and the conclusions of \citet{Numata15}, who performed gyrokinetic simulations (implicitly assuming strong guide fields) of non-relativistic reconnection. They found an excess of electron heating at low beta, similar to the qualitative predictions of the \citet{Howes10} prescription.\footnote{Still, the ratio of ion to electron heating efficiency that they measured at low beta ($\sim 10^{-3}$ for $\beta_{\rm i}=0.01$) is much higher than the prediction of the \citet{Howes10} model.} 
A further fundamental difference between might lie in the fact that the simulations by \citet{Numata15} at $\beta_{\rm i}=0.01$ did not show any secondary plasmoids, possibly as a consequence of the limited domain size, whereas the reconnection layer in \citet{Rowan17}'s PIC simulations at low $\beta_{\rm i}$ is copiously fragmented into secondary magnetic islands (fig. 5 of \citealt{Rowan17}). Such secondary plasmoids are efficient sites of ion heating. 
The efficiency of electron heating in PIC simulations of plasmoid-dominated reconnection in the strong guide field regime will be presented in a forthcoming work. Furthermore, we note that reconnection in collisionless accretion flows itself likely occurs at the endpoint of a turbulent cascade. Considering the effects of turbulence  between the grid scale and the reconnection scale may modify the results used here (see e.g. \citealt{Shay2018}).

Finally, we note that two-fluid studies of the magnetorotational instability (MRI)  using shearing box simulations have shown that pressure anisotropies can lead to substantial viscous heating at large scales \citep[e.g.][]{Sharma2007,Riquelme2012,NarayanSironi,Sironi15, Riquelme2016}. \citet{Sharma2007} note that this large-scale dissipation could account for 50 per cent of the electron heating in systems like \sgraa. 
This large-scale heating can be captured in a two-fluid GRMHD simulation that includes viscous stresses from pressure anisotropy. \citet{Chandra15} introduced a formalism later implemented in the code \texttt{grim} \citep{Chandra17} which captures viscous heating from pressure anisotropies in a single-fluid framework. However, while \citet{Ressler15,Ressler17} include anisotropic electron heat flux along field lines in their two-temperature simulations, they do not include the pressure anisotropy term that could give rise to large-scale electron heating. We also neglect this large-scale heating in the present work and focus exclusively on grid-scale dissipation.

\section{Numerical Simulations}
\label{sec::num}

\begin{table*}
\centering
\caption{
Simulation Parameters
}
\label{tab::summary}
\begin{tabular}{l||llll|llll}
\hline \hline
Model  & Spin & Heating & $r_\text{min}$ & $r_\text{max}$ & $N_r \,\times\, N_\theta \,\times\, N_\phi$ & $[t_\text{i},t_\text{f}]\;(t_{\rm g})$ & $\dot{M} (\dot{M}_\mathrm{Edd})$ & $\Phi_{\rm BH} \; \left((\dot{M}c)^{1/2}r_{\rm g}\right)$ \\
\hline
H-Lo   &  0      & Turb. Cascade  & 1.5 & 5000 & $320\times192\times96$ & $[2.8\times10^4,\;3.3\times10^4]$ & $3\times10^{-7}$ & 5  \\
R-Lo   &  0      & Mag. Reconnection  & 1.5 & 5000 & $320\times192\times96$ & $[2.5\times10^4,\;3.0\times10^4]$ & $7\times10^{-7}$ & 4  \\
H-Hi   &  0.9375 & Turb. Cascade  & 1   & 5000 & $320\times192\times96$ & $[2.7\times10^4,\;3.2\times10^4]$ & $2\times10^{-7}$ & 6  \\
R-Hi   &  0.9375 & Mag. Reconnection  & 1   & 5000 & $320\times192\times96$ & $[2.8\times10^4,\;3.3\times10^4]$ & $3\times10^{-7}$ & 3  
\end{tabular}
\end{table*}

\subsection{Units}
\label{sec::units}

In the accretion simulations presented below, we fix the  
black hole mass to a value appropriate for \sgraa, $M=4\times10^6 \, M_\odot$
\citep{Gillessen09,Chatzopoulos15}. We use the gravitational
radius $r_{\rm g} = GM/c^2 = 6\times10^{11} \,\text{cm} = 0.04 \,\text{AU}$
as the unit of length, and $t_{\rm g} = r_{\rm g}/c = 20$\,s as the unit of time. 
The Eddington accretion rate is defined as
\begin{equation}
 \dot{M}_\mathrm{Edd} = \frac{L_\mathrm{Edd}}{\eta c^2} = \frac{4\pi \, GMm_p}{\eta c \, 
\sigma_{\rm T}},
\end{equation}
where $L_\mathrm{Edd}$ is the Eddington luminosity, and we set the
efficiency $\eta = 0.057$ for both spins. Thus for \sgraa, the
Eddington accretion rate $\dot{M}_\mathrm{Edd}=0.16 \, M_\odot$
yr$^{-1}$, and the Eddington luminosity
$L_\mathrm{Edd}~=~5\times10^{44}$ erg s$^{-1}$.

\subsection{Simulation Grid and Initial Conditions}
\label{sec::init}

We set up our simulations in the Kerr metric using a modified Kerr-Schild coordinate grid.
The central black hole in each model had a mass of $4 \times 10^6 M_\odot$. We used two spin values 
in our simulations: $a=0$ (spin zero case) and $a=0.9375$ (high spin case). At both  spins we
ran two simulations, one for each of the heating prescriptions described in Section~\ref{sec::heating}. Thus, we have  four models: 
a spin zero turbulent heating prescription model, H-Lo, a spin zero model using our new magnetic reconnection heating prescription, R-Lo, 
a spin 0.935 turbulent heating prescription model, H-Hi, and a corresponding 0.9375 spin model heated by magnetic reconnection, R-Hi.
The simulation parameters are summarized in Table~\ref{tab::summary}.

We ran each simulation in a 3D grid with a resolution of $256\times192\times96$ cells in
radius, polar angle, and azimuth. The radial cells are distributed exponentially
from a Boyer-Lindquist radius $r_\text{min}$ inside the black hole horizon out to $5\times10^3 \, r_{\rm g}$. The azimuthal cells are distributed uniformly over the range $[-\pi, \pi]$, while the polar angle cells are sampled over the interval $[0,\pi]$ using the function presented in Appendix~\ref{appendix::grid}. To better resolve the MRI in the disc, our grids more densely sample the regions closer to the equatorial plane. We also follow \citet{Tchekhovskoy11} in `cylindrifying' the polar axis cells near the black hole; this speeds up the simulation by increasing the cell sizes along the axis near the black hole, relaxing the global Courant limit on the time step size. 

The initial gas torii were set up using the \citet{Penna13} model, with weak dipolar field loops added in the $(r,\theta)$ plane. The initial energy in electrons was taken as one per cent of the total gas energy, with the remainder in ions. Our initial torii and simulation grids are presented in  more detail in Appendix~\ref{appendix::grid}, and displayed in Fig.~\ref{fig::init}.

To slightly speed up our simulations during the initial stages, we first ran the models for a total time of $2\times10^4 \, t_{\rm g}$ in 2D, suppressing the $\phi$ coordinate and assuming axisymmetry. During this step we used the mean-field dynamo presented in \citet{KORAL15} to prevent the decay of the magnetic field. After running the simulations for $2\times10^4\,t_{\rm g}$ in 2D, we regridded the output into 3D, rescaled the density by a factor of 10,  and introduced 5 per cent perturbations in the azimuthal three-velocity $v_\phi$ to seed departures from axisymmetry. The rescaling factor  of 10 was chosen from test simulations in order to achieve the desired accretion rate ($\dot{M} \sim 10^{-7}\,\dot{M}_{\rm Edd}$) in the 3D run; since the level of angular momentum transport 
facilitated by the self-consistent MRI turbulence is somewhat less than that supplied by the dynamo in 2D, the accretion rate was generally lower in 3D than in 2D for the same gas density. We ran the simulations in 3D for another $1.5\times10^4 \, t_{\rm g}$, with the mean-field dynamo turned off. 

In evolving the explicit part of the GRRMHD equations and the electron/ion thermodynamic equations, we used the piecewise parabolic method (PPM) to reconstruct fluxes at the cell walls. We found that using PPM produced results in 3D with more stable accretion rates over the $5000$ to $10{,}000\,r_{\rm g}$ intervals we consider than the linear flux-limiter methods often used in GRMHD simulations. 
 
Throughout the run of each simulation, we saved snapshots of the evolved variables every $10 \, t_{\rm g}$. The results presented below correspond
to a $5000 \, t_{\rm g}$ period for each simulation, taken from somewhere between $2.5\times10^4 t_{\rm g}$ and $3.5 \times 10^4 t_{\rm g}$ (see Table~\ref{tab::summary} for the exact range selected for each simulation).

\subsection{Radiative Transfer}
We compute spectra and lightcurves from our simulations using the post-processing code 
\texttt{HEROIC}, \citep{Zhu_2015, Narayan16_HEROIC}. 
\texttt{HEROIC} solves for a spectrum and angular distribution of radiation 
at each grid position self-consistently using the geodesic equation
and radiative transfer equation. This self-consistent solution allows 
different geodesics to exchange intensities due to scattering of photons by electrons.
We include synchrotron, bremsstrahlung, and inverse Compton scattering in the \texttt{HEROIC} post-processing. At the 1.3~mm observing wavelength of the EHT, synchrotron radiation dominates the emission.
To produce higher-resolution images of the
accretion flow at 1.3~mm wavelength, we use the ray tracing and radiative transfer code \texttt{grtrans} \citep{Dexter16} using only thermal synchrotron opacities. 

We compute spectra and lightcurves for several different inclination angles 
($10^\circ$, $20^\circ$, $40^\circ$, $60^\circ$, $80^\circ$, and $90^\circ$), 
measured down from the north pole. In computing lightcurves, we use the `fast light'
approximation, where individual images
are computed using fixed lab frame time slices of the simulation output, not allowing
allow the fluid to evolve as photons propagate in postprocessing.

Before running \texttt{HEROIC}, we scale the density and magnetic field in 
the simulations (keeping the electron temperature fixed), so as to match the average observed \sgraa\ 1.3~mm flux density ($\approx 3.5$ Jy: \citealt{Bower2015}) at $60^\circ$ inclination. The scaling factors were different for each model, ranging from 0.06 (model H-Hi) to 1.75 (model R-Lo). Because our two-temperature GRRMHD simulations include radiation and Coulomb couplings that are not scale-free, this procedure is not consistent if these couplings are dynamically important. For \sgraa\, Coulomb and radiation couplings do not significantly alter the gas dynamics, so a limited amount of rescaling in post-processing should not affect the validity of our results. A more consistent procedure would be to  identify rescaling factors and then re-run the selected part of the simulation with the scaled primitives. This will be particularly important in higher accretion rate systems where radiation coupling starts to become important, such  as M87 \citep{Ryan18}. 

Because $\sigma_\mathrm{i}$ can exceed unity in the jet region close to the poles,
the plasma dynamics in this region are  dominated by the magnetic field.
Small errors in the total conserved energy in the simulation 
can then lead to large errors in the fluid energy and hence
the electron temperature. Furthermore, the density is extremely 
low in this region and is often determined by an arbitrary floor. For these reasons,
we remove the innermost 4 layers of cells closest to each polar axis in our postprocessing. 

\section{Results}
\label{sec::results}

\subsection{Accretion Flow Properties}
\begin{figure*}
\centering
\includegraphics*[width=0.99\textwidth]{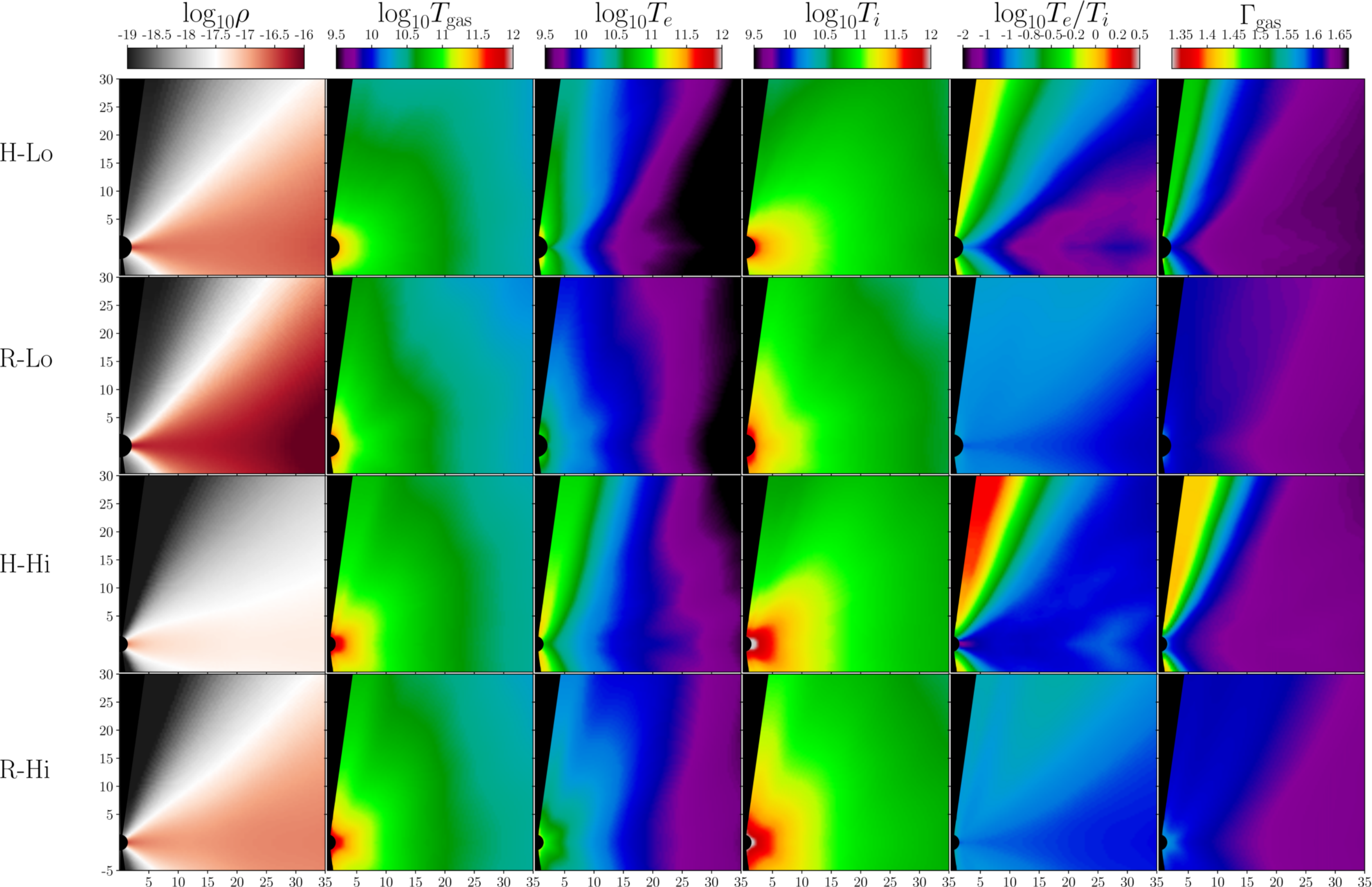}
\caption{Bulk gas properties of the four simulations. From top to bottom, quantities are shown for the spin 0 turbulent heating model H-Lo, the spin 0 reconnection  model R-Lo, the spin 0.9375 turbulent heating model H-Hi, and the spin 0.9375 reconnection model R-Hi. The fluid quantities were rescaled to produce a 230~GHz flux density around 3.5~Jy \citep{Bower2015} when observed at $60^\circ$ inclination, and were averaged in azimuth and time for $5000 \, t_{\rm g}$. The resulting averages were symmetrized over the equatorial plane. 
From left to right, the quantities displayed are the density $\rho$
in g cm$^{-3}$, the gas temperature $T_\mathrm{gas}$ in K, the electron temperature $T_\mathrm{e}$ in K, the ion temperature $T_\mathrm{i}$ in K, the electron-to-ion temperature ratio $T_\mathrm{e}/T_\mathrm{i}$, and the effective gas adiabatic index $\Gamma_\mathrm{gas}$.
}
\label{fig::symcompare}
\end{figure*}

\begin{figure*}
\centering
\includegraphics*[width=0.99\textwidth]{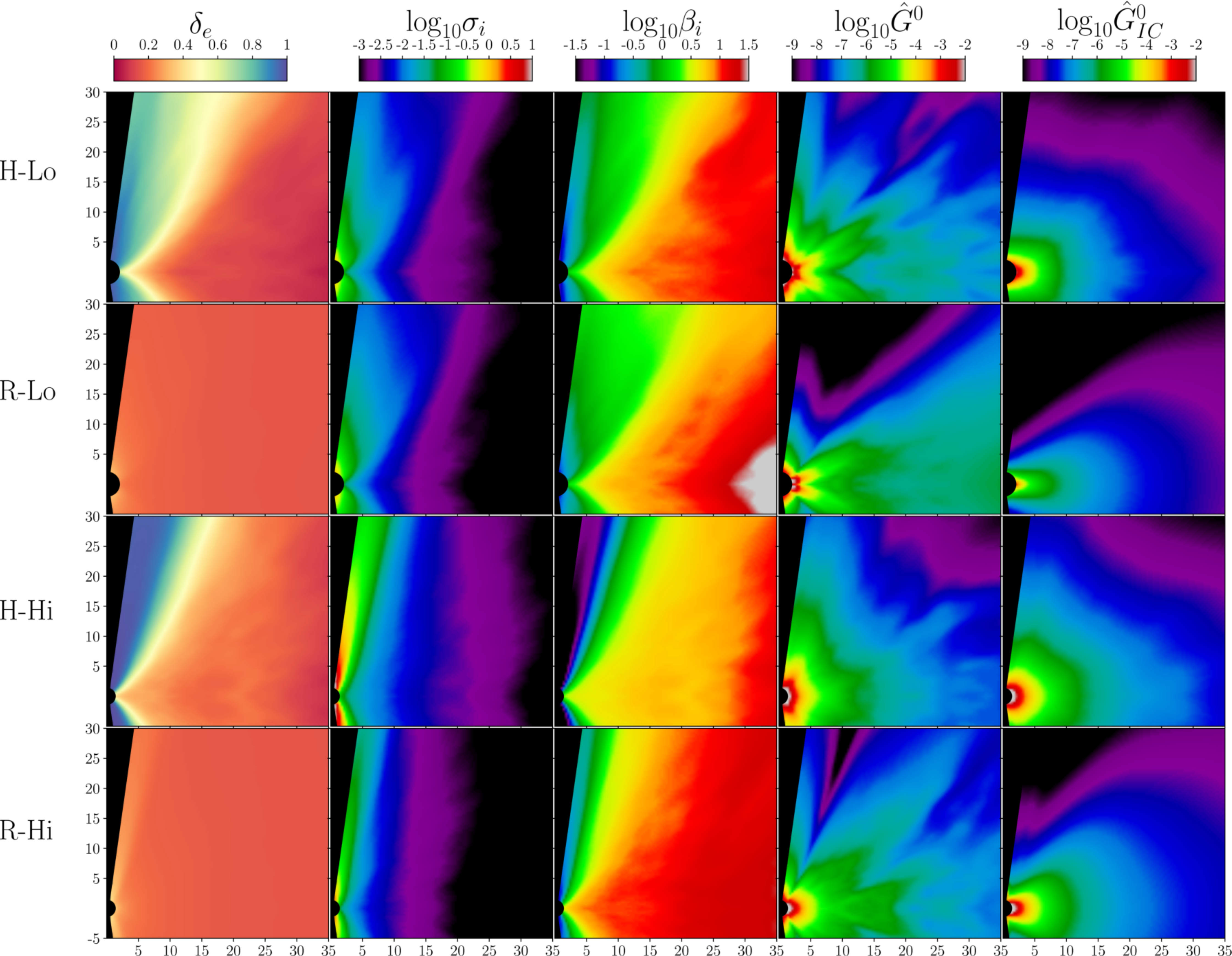}
\caption{Additional azimuth and time-averaged properties of the four models. 
From left to right, the quantities displayed are the electron heating fraction $\delta_\mathrm{e}$,
the plasma magnetization $\sigma_\mathrm{i}$, 
the ratio of ion thermal pressure to magnetic pressure $\beta_\mathrm{i}$,
the total rest frame radiation power $\hat{G}^0$ in erg s$^{-1}$ cm$^{-3}$, and the inverse Compton
radiation  power $\hat{G}^0_{IC}$ in the same units.
}
\label{fig::symcompare2}
\end{figure*}

\begin{figure*}
\centering
\includegraphics*[width=0.9\textwidth]{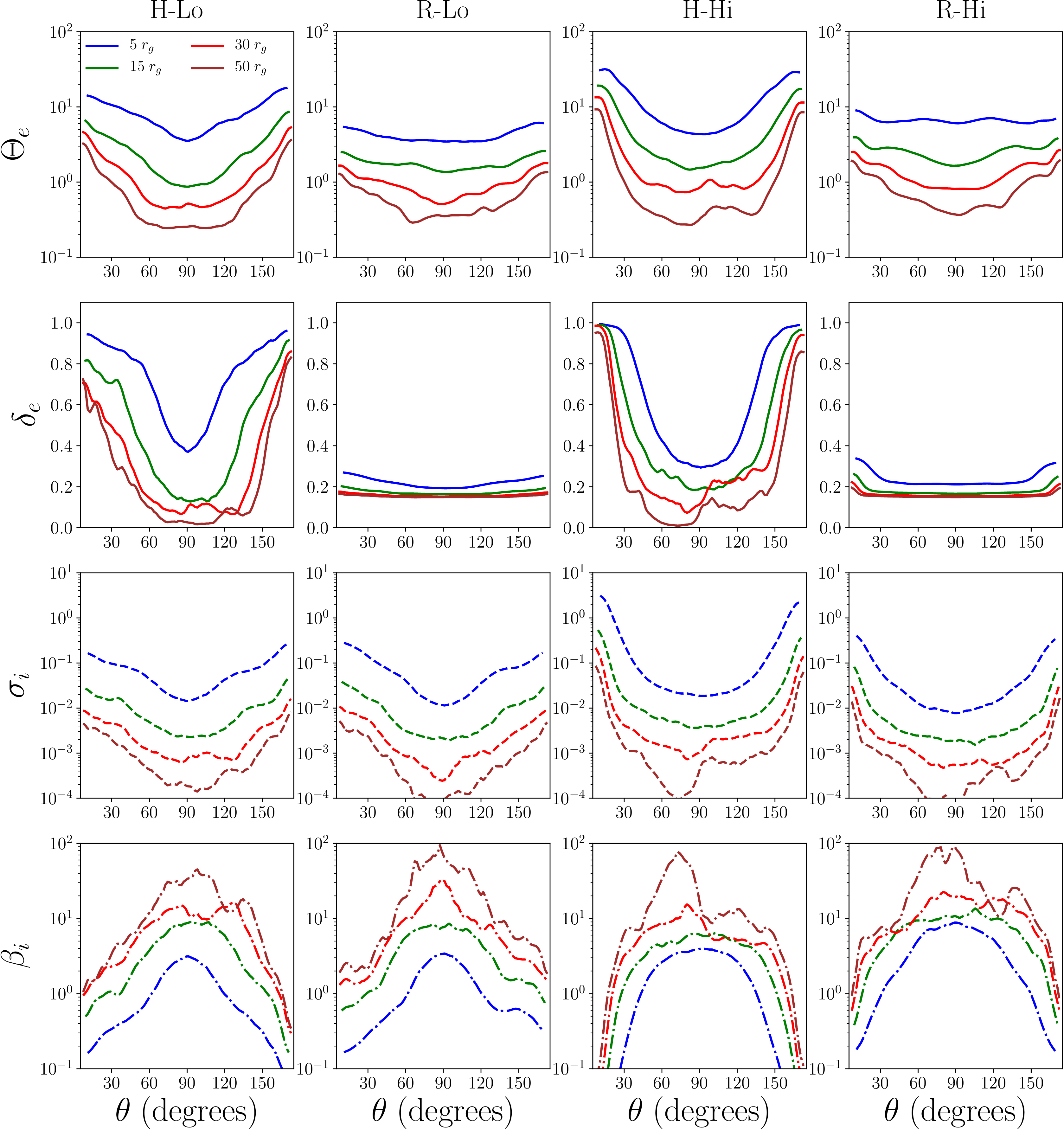}
\caption{
Azimuth and time-averaged fluid properties as a function of polar angle $\theta$ at four radii.
Unlike in Figs.~\ref{fig::symcompare} and~\ref{fig::symcompare2}, these data were not symmetrized over the equatorial plane.
All quantities are plotted for each model at radii $5 \, r_{\rm g}$ (blue), $15 \, r_{\rm g}$ (green),  $30 \, r_{\rm g}$ (red), and $50 \, r_{\rm g}$ (brown).
From top to bottom, the quantities displayed are the dimensionless electron temperature $\Theta_\mathrm{e} = k_{\rm B} T_\mathrm{e} / m_\mathrm{e} c^2$,  
the electron heating fraction $\delta_\mathrm{e}$,
the plasma magnetization $\sigma_\mathrm{i}$, and
the ratio of ion thermal pressure to magnetic pressure $\beta_\mathrm{i}$.
}
\label{fig::symcompare_curves}
\end{figure*}

Figs.~\ref{fig::symcompare},~\ref{fig::symcompare2}, and~\ref{fig::symcompare_curves} show quantities averaged in azimuth and time over the $5000 \, t_{\rm g}$ period selected for each model (see Table~\ref{tab::summary}). Fig.~\ref{fig::symcompare} displays the mass density $\rho$, the gas temperature $T_\mathrm{gas}$,  the electron temperature $T_\mathrm{e}$, the ion temperature $T_\mathrm{i}$, the temperature ratio $T_\mathrm{e}/T_\mathrm{i}$, and the effective gas adiabatic index, $\Gamma_\mathrm{gas}$ (see equation~\ref{eq::gammaeff}). Fig.~\ref{fig::symcompare2} shows the electron heating fraction $\delta_\mathrm{e}$, the magnetization $\sigma_\mathrm{i}$, the ratio of ion thermal pressure to magnetic pressure $\beta_\mathrm{i}$, the fluid frame radiation power per unit volume $\hat{G}^0$, and the power per unit volume produced by inverse Compton scattering $\hat{G}^0_{IC}$, as calculated in KORAL using frequency-averaged quantities \citep{Compt15}. Fig.~\ref{fig::symcompare_curves} shows angular profiles of the dimensionless electron temperature, $\Theta_\mathrm{e}=k_{\rm B} T_\mathrm{e}/m_\mathrm{e} c^2$, $\delta_\mathrm{e}$, $\sigma_\mathrm{i}$, and $\beta_\mathrm{i}$, taken at three different radii. Before averaging, the primitive quantities in each simulation were scaled so as to produce an average 230~GHz flux density of approximately 3.5 Jy. Derived quantities like the species temperatures and radiation power were then recomputed from the scaled primitives.  Temperatures and dimensionless ratios are unchanged by this scaling, but the density and radiation power profiles are affected. 

In Fig.~\ref{fig::symcompare}, we see that all four models produce discs that are geometrically thick. Each simulation was run from an initial torus with almost identical initial density profiles with only a slight difference between the two spins considered. However, because of the rescaling required to produce a 230~GHz flux density near the measured value for \sgraa, the final density profiles show significant differences among the models. In particular, due to cooler electron temperatures in the funnel and inner disc, the density in the low spin magnetic reconnection heating model R-Lo had to be scaled up to produce the right 230~GHz flux density. In contrast, the high electron temperatures in the jet/funnel region and stronger magnetic field in model H-Hi required a large downscaling in density. These differences in density are also apparent in the average accretion rates presented in Table~\ref{tab::summary}; disc R-Lo has the highest accretion rate of $7\times10^{-7} \,\dot{M}_\mathrm{Edd}$, while disc H-Hi has the lowest overall accretion rate of $2\times10^{-7} \,\dot{M}_\mathrm{Edd}.$ The  accretion rates for all four models fall within the limits for \sgraa\ set by Faraday rotation measurements \citep{Marrone_2007}. 

In addition, the $\beta_\mathrm{i}$ and $\sigma_\mathrm{i}$ distributions for the models plotted in Figs.~\ref{fig::symcompare2} and~\ref{fig::symcompare_curves} show that all four models are largely in the same regime with regards to pressure and magnetic field strength, with some notable differences. All four models have low levels of magnetic flux, with the amount of time-averaged flux threading the black hole $\Phi /(\dot{M}c)^{1/2}r_{\rm g} < 10$ in all cases. This puts our discs squarely in the Standard and Normal Evolution regime of accretion \citep[SANE;][]{Narayan2012}, well below a Magnetically Arrested Disc \citep[MAD;][]{Bisnovatyi1976, NarayanMAD, Igumenschchev2003, Tchekhovskoy11} regime, where $\Phi_{\rm BH}/(\dot{M}c)^{1/2}r_\mathrm{g} \approx50$. The two low-spin models have lower field strengths, with $\sigma_\mathrm{i}<0.1$ everywhere except extremely close to the black hole, and $\sigma_\mathrm{i}$ falls to $10^{-3}$ past $20\,r_{\rm g}$. Both high spin models produce more field strength in the jet region, but $\sigma_\mathrm{i}$ still falls rapidly with radius in the equatorial plane.  Of the four models, the turbulent heating prescription at high spin, H-Hi, has the most magnetic flux. H-Hi achieves $\sigma_\mathrm{i}\sim1$ in the jet region close to the black hole, and has higher values of $\sigma_\mathrm{i}$ (and lower values of $\beta_\mathrm{i}$) at all radii compared to the magnetic reconnection model at the same spin. This model also launches a mildly relativistic jet, with Lorentz factor $\approx2$ at large radii. 

While Fig.~\ref{fig::symcompare} shows that the gas temperature distributions are similar in all four models, the electron and ion temperatures vary dramatically with the choice of heating prescription. The distribution of the electron heating fraction $\delta_\mathrm{e}$ in the simulations is distinct for each heating prescription, but shows only slight differences with spin. Because the turbulent heating prescription deposits most of the dissipated energy into electrons at low $\beta_\mathrm{i}$, models H-Lo and H-Hi both show higher values of $\delta_\mathrm{e}>0.5$ in the funnel. The more magnetized jet in model H-Hi makes this transition sharper, and results in $\delta_\mathrm{e}\sim1$ at low polar angles for  all radii (most easily seen in the angular profiles in Fig.~\ref{fig::symcompare_curves}), while in model H-Lo, $\delta_\mathrm{e}$  only approaches unity at small radius. This distribution of $\delta_\mathrm{e}$ produces electrons that are hotter in the jet/funnel, consistent with the simulations reported by \citet{Ressler17} and \citet{KORAL16}, though the absolute temperatures seen in our models are lower due to the weaker magnetic field. In all our models, electrons are relativistic in the outflow and inner disc, with $\Theta_\mathrm{e}>1$ ($T_\mathrm{e} >6\times 10^9$ K), but we do not observe the very high electron temperatures $\Theta_\mathrm{e}\sim100$ seen in more magnetically dominated simulations. Fig.~\ref{fig::symcompare} shows that the temperature ratio $T_\mathrm{e}/T_\mathrm{i}$ takes on values between 0.05 and 3 for the turbulent heating prescription, with an obvious structure proceeding from $T_\mathrm{e}<T_\mathrm{i}$  in the disc near the equator to $T_\mathrm{e}>T_\mathrm{i}$ in the outflow and jet regions where the magnetic field strength is larger, $\beta_\mathrm{i}$ is lower, and $\delta_\mathrm{e} \rightarrow 1$. 

In contrast, the magnetic reconnection prescription never heats the electrons more than ions. Fig.~\ref{fig::symcompare_curves} shows that in the weakly magnetized regime explored by our models, $\delta_\mathrm{e}$ varies little with polar angle and does not exceed 0.3 on average. As a result, we find $T_\mathrm{e}/T_\mathrm{i}<1$ everywhere. While the magnetic reconnection fitting function does put more heat in the electrons outside of the midplane where $\beta_\mathrm{i}$ is lower, the effect is small. $T_\mathrm{e}/T_\mathrm{i}$ takes on a value around 0.1 near the equator and climbs to only around 0.15 at larger polar angles. The electron temperatures in the outer disc, relevant for free-free X-ray emission, are similar in both models (around $10^9$ K), despite the slightly larger electron heating delivered by the reconnection  model at these radii (see the $50 \, r_{\rm g}$ curve in Fig.~\ref{fig::symcompare_curves}). 

The last column of Fig.~\ref{fig::symcompare} shows the effects of the electron heating on the total gas adiabatic index (equation~\ref{eq::gammaeff}). Both turbulent heating models H-Lo and H-Hi show the total gas adiabatic index $\Gamma_\mathrm{gas}$ dropping to $\approx 1.4$ in the funnel, as in this region relativistic electrons with $\Gamma_\mathrm{e} \approx 4/3$ start to dominate the fluid's energy budget. However, even in the models heated by magnetic reconnection, where electrons always have less than 50 per cent of the total gas energy, the gas adiabatic index is not exactly $\Gamma_\mathrm{gas} = 5/3$. Out to $\approx20 \, r_{\rm g}$, the adiabatic index is closer to 1.6 than exactly $5/3$, indicating the effects of relativistic electrons in lowering the total gas adiabatic index even when they do not constitute a majority of the fluid energy. 

As a result of the distinct electron temperature distributions that result from the two heating prescriptions, the distributions of radiation power in Fig.~\ref{fig::symcompare2} also are different. In the turbulent heating prescription at low spin, H-Lo, high electron temperatures in the outflow result in a bolometric radiation power distribution (both in synchrotron and inverse Compton) that is roughly spherical. This spherical distribution is present both in the total radiation power $\hat{G}^0$, which is dominated by synchrotron emission, and in the inverse Compton power $\hat{G}^0_{IC}$. In the magnetic reconnection heating model R-Lo, electrons in the outflow are not dramatically hotter than electrons in the disc, and due to their low density make only a small contribution to the overall radiation power. As a result, the distributions of synchrotron and inverse Compton power in these models are disc-dominated. At high spin the picture remains largely the same. Hotter electron temperatures close to the black hole cause the magnetic reconnection model R-Hi to  have a more isotropic distribution of synchrotron power, but the contribution from the jet region is still significantly less than in the high spin turbulent heating model, H-Hi. At both low and high spin, the distributions of inverse Compton power show that using the magnetic reconnection heating model (R-Lo, R-Hi), the Compton scattering is confined to the disc, whereas with the turbulent heating model (H-Lo, H-Hi), inverse Compton produces significant power at all angles. 

\subsection{Spectra}
\label{sec::spectra}
\begin{figure*}
\centering
\includegraphics*[width=0.99\textwidth]{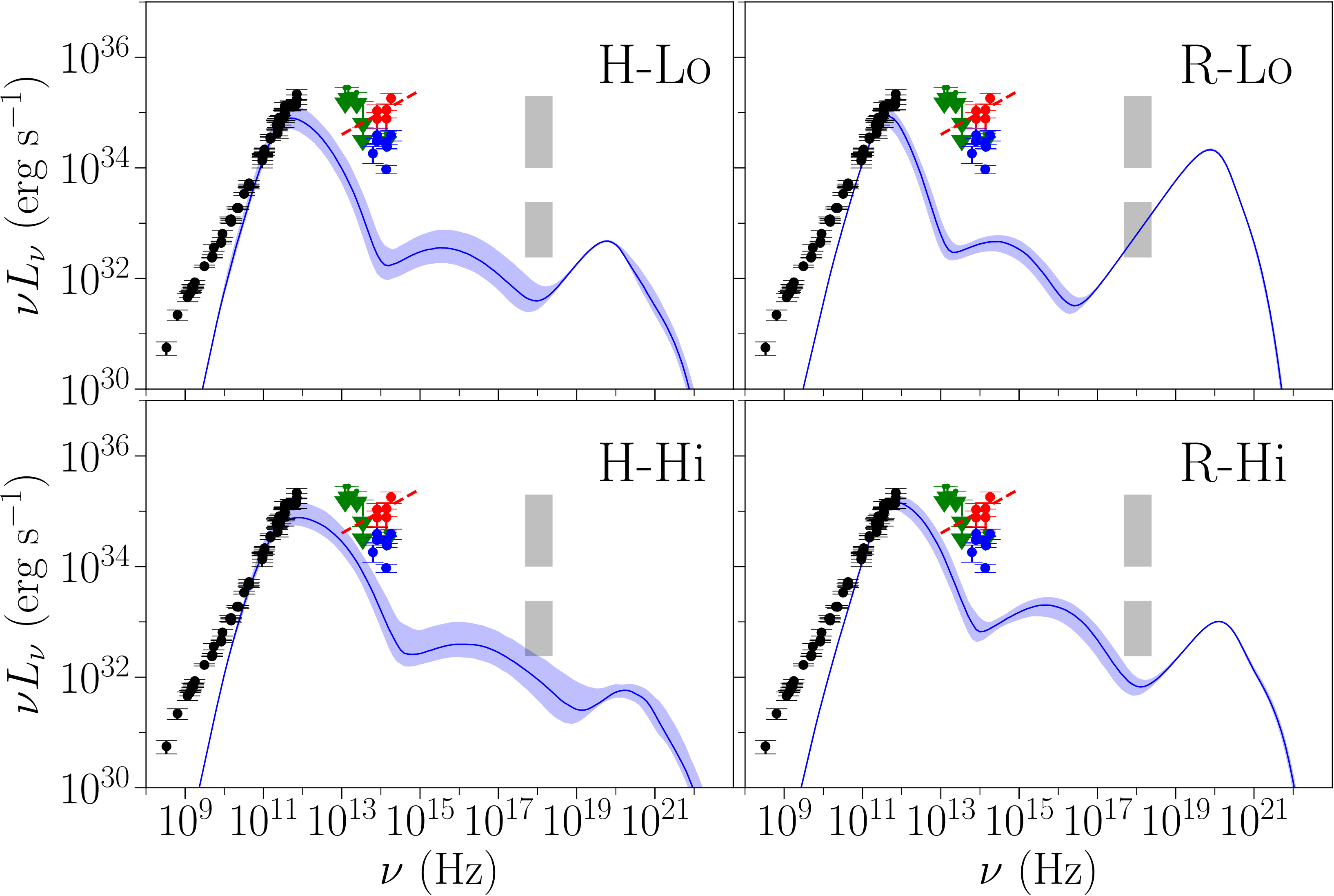}
\caption{
Spectral energy distributions for the four models, calculated with \texttt{HEROIC} for an observer at $60^\circ$ inclination. Spectra were computed every $10 \, t_{\rm g}$ over a $5000 \,t_{\rm g}$ period (see Table~\ref{tab::summary} for the exact ranges). The solid blue curve shows the median spectrum for each model, and the shaded blue region shows the nominal $1\sigma$ time-variability if we assume that the variability distribution is Gaussian at each frequency. Data points in the radio and near-infrared are taken from references listed in Appendix~\ref{appendix::data}. Black data points show radio measurements. Green data points are near-infrared upper limits, blue data points are near-infrared quiescent measurements, and red data points are near-infrared flare measurements.  The near-infrared spectral slope shown by the red line was taken from the flare measurement in \citet{Gillessen2006} as $\nu L_\nu \propto \nu^{0.4}$. The lower shaded vertical band in the X-ray represents the range of potential X-ray quiescent emission from the inner region of \sgraa, between 10 and 100 per cent of the total observed \citep{Baganoff03}. The upper shaded vertical X-ray band shows the range of observed X-ray flares \citep{Neilsen}. From left to right, top to bottom, spectra are shown for the spin 0 turbulent heating model H-Lo, the spin 0 magnetic reconnection  model R-Lo, the spin 0.9375 turbulent heating model R-Hi, and the spin 0.9375 reconnection  model R-Hi. 
}
\label{fig::spectra}
\end{figure*}

\begin{figure}
\centering
\includegraphics*[width=0.35\textwidth]{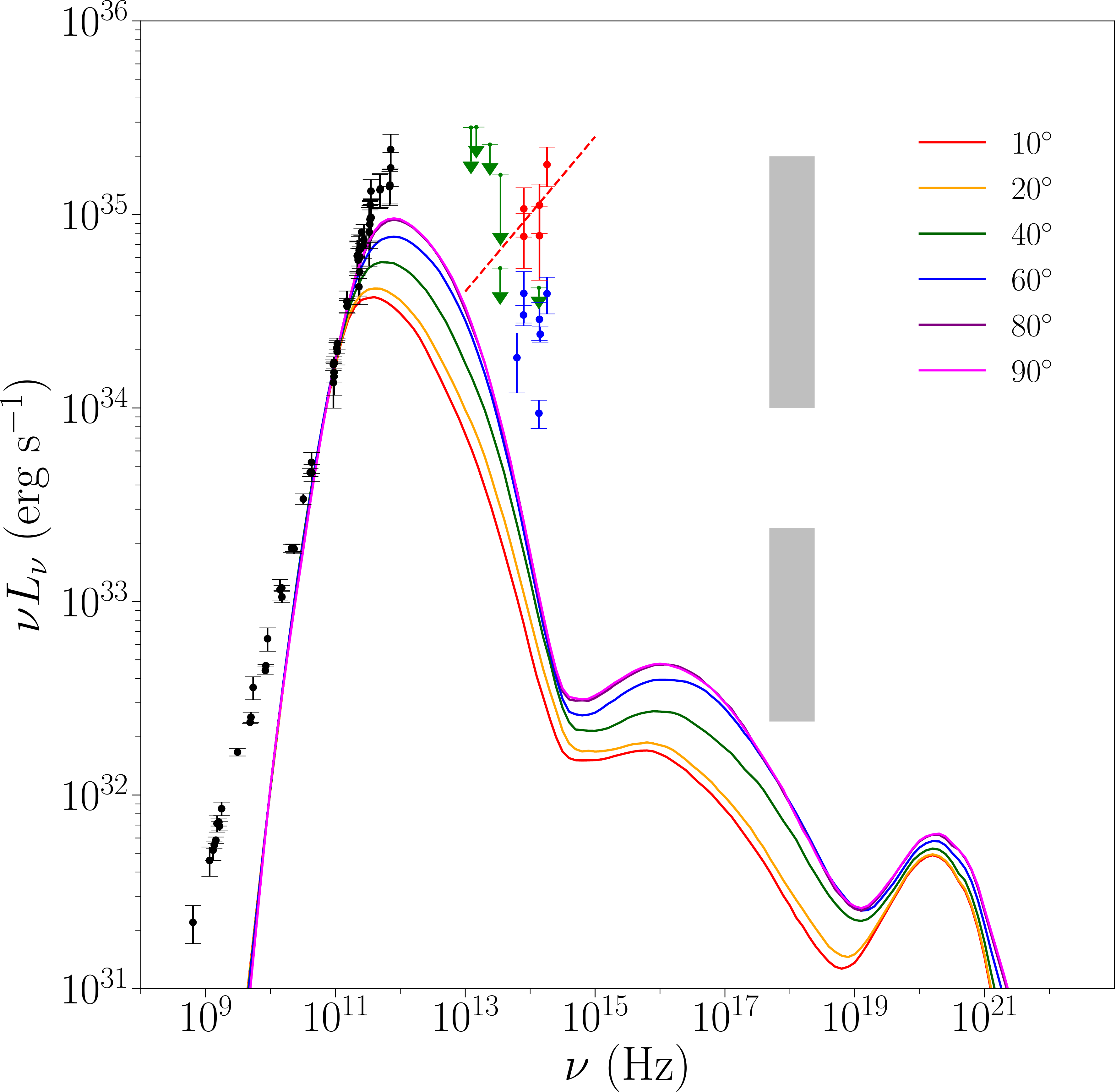}
\caption{
Median spectra for the model H-Hi taken at different observer inclinations. Due to relativistic beaming of material orbiting very close to the spin 0.9375 black hole, the main synchrotron peak of the spectrum becomes more intense when the system is viewed edge-on, 
while the free-free X-ray emission from further out in the disc does not change appreciably with inclination.
}
\label{fig::spectra_inc}
\end{figure}

After rescaling the density so that the average 230~GHz flux density is approximately equal to that observed from  \sgraa\ (3.5 Jy), we computed spectra with \texttt{HEROIC} at a cadence of every $10 \, t_{\rm g}$ within the $5000 \, t_{\rm g}$ window selected for each model. In Fig.~\ref{fig::spectra} we display the median Spectral Energy Distributions (SEDs) for all four models observed at $60^\circ$ inclination, with the nominal $\pm 1\sigma$ variability around the median denoted by the shaded region (assuming the spectrum at each frequency is Gaussian distributed, we take as the $\pm 1\sigma$ band the 68 per cent interval between the 15.9th percentile and the 84.1th percentile). All models show the same characteristic features. At frequencies lower than $10^{11}$ Hz, the spectrum is dominated by  optically thick synchrotron emission from the outer disc and outflow. The spectrum transitions to an optically thin synchrotron peak around $10^{12}$ Hz produced by emission in the inner disc close to the black hole. An inverse Compton hump between $10^{14}$ and $10^{18}$ Hz is produced from the Compton upscattering of NIR photons, and at X-ray frequencies the emission is dominated by thermal free-free emission from radii out to $r\sim50 r_{\rm g}$, peaking at $10^{20}$ Hz. Fig.~\ref{fig::spectra_inc} shows the effect of inclination angle on the spectrum. Relativistic beaming from viewing the accretion disc edge-on ($90^\circ$) increases the luminosity of the synchrotron peak and Compton hump from scattered synchrotron photons, but does not significantly affect the free-free X-ray emission produced at larger radii. This effect is more pronounced for the spin 0.9375 models, as the orbital  velocity in these models is close to the black hole allows for more intense relativistic beaming from higher-velocity gas. 

The inverse Compton hump of the turbulent heating models (H-Lo and H-Hi) is at higher frequencies compared to the models heated by magnetic reconnection (R-Lo and R-Hi). This is due to the fact that in the H models, inverse Compton emission is produced in a spherical region around the black hole, including contributions from hotter electrons in the outflow, whereas in the R models, inverse Compton scattering is confined to  cooler electrons in the disc (see Fig.~\ref{fig::symcompare2}). 

All models match measurements of the high-frequency radio spectrum from optically thin synchrotron between $10^{10.5}$ and $10^{12}$ Hz reasonably well. However, all models under-predict the spectrum at low frequencies $\nu<10^{10.5}$ Hz. The relatively flat low frequency slope ($L_\nu \propto \nu^{0.2}$) \citep{Falcke98,Hern04} could be the result of an isothermal jet or outflow not captured by our simulation \citep{Moscibrodzka_13, Ressler17}, or by emission from a population of high-energy non-thermal electrons \citep{Ozel2000, Yuan2003, Davelaar_17}. We note that Model H-Hi does the best job of fitting the low frequency spectrum down to $\sim10^{10.5}$~Hz whereas the other three models start failing to fit the data around $10^{11}$~Hz. This may be due to the jet emission which dominates the H-Hi model images at low frequencies (see Section~\ref{sec::images}). Consequently, the inverse Compton emission in models H-Lo and H-Hi is also more variable than in the corresponding magnetic reconnection models R-Lo and R-Hi. 

X-ray emission is primarily produced by thermal bremsstrahlung at larger radii out to $r\sim50 \, r_{\rm g}$, the largest radius we use in our radiative transfer calculations. Because all four models have roughly similar electron temperatures $\sim10^9$ K, the strength of the free-free peak at around $10^{20}$~Hz is thus primarily determined by the disc density around this radius, which is in turn set by the rescaling factor chosen to match the observed 230~GHz flux density. At each spin, because the turbulent heating models (H-Lo and H-Hi) produce higher electron temperatures in the inner disc and outflow, they are scaled to a lower density than the corresponding magnetic reconnection heating models (R-Lo and R-Hi). Similarly, the high spin models produce hotter electron temperatures close to the black hole and therefore have density scaling factors smaller than the corresponding spin zero models, thus lowering their thermal free-free X-ray peaks. With the exception of model R-Lo, all of the X-ray spectra lie below the estimated quiescent luminosity from the inner disc, which we take as 10 per cent of the total 2-10 keV X-ray luminosity measured by \citet{Baganoff03}. With the addition of free-free emission outside the maximum radius of $r=50 \, r_{\rm g}$ used in our radiative transfer, it is likely that model R-Lo would exceed the total luminosity measured by \citet{Baganoff03}.

All models substantially under-predict the observed quiescent near-infrared emission (blue data points in Fig.~\ref{fig::spectra}), and the observed variability does not produce flares as strong as those observed in \sgraa\ (the red data points in Fig.~\ref{fig::spectra}). In addition, the near-infrared spectral slope in our models is sharply negative, whereas the spectral slope measured in near-infrared flares is positive ($\nu L_\nu \propto \nu^{0.4}$,
\citealt{Genzel03,Gillessen2006,Hornstein2007}). 

The positive near-infrared spectral index suggests that flares may be produced by non-thermal electrons which are not considered in our simulations. \citet{Ponti2017} measured the spectral index of a single strong flare in the near-infrared and X-ray. In addition to confirming $\nu L_\nu \propto \nu^{0.4}$ in near-infrared, they found the  found a difference of $\approx 0.5$ between the X-ray and near-infrared spectral indices, suggestive of power-law emission with a synchrotron cooling break. In future work using the method of \citet{Chael17}, we will explore the production of near-infrared synchrotron emission from electrons accelerated into power-law distributions around \sgraa. 

\subsection{Variability}
\label{sec::variability}
\begin{figure*}
\centering
\includegraphics*[width=0.95\textwidth]{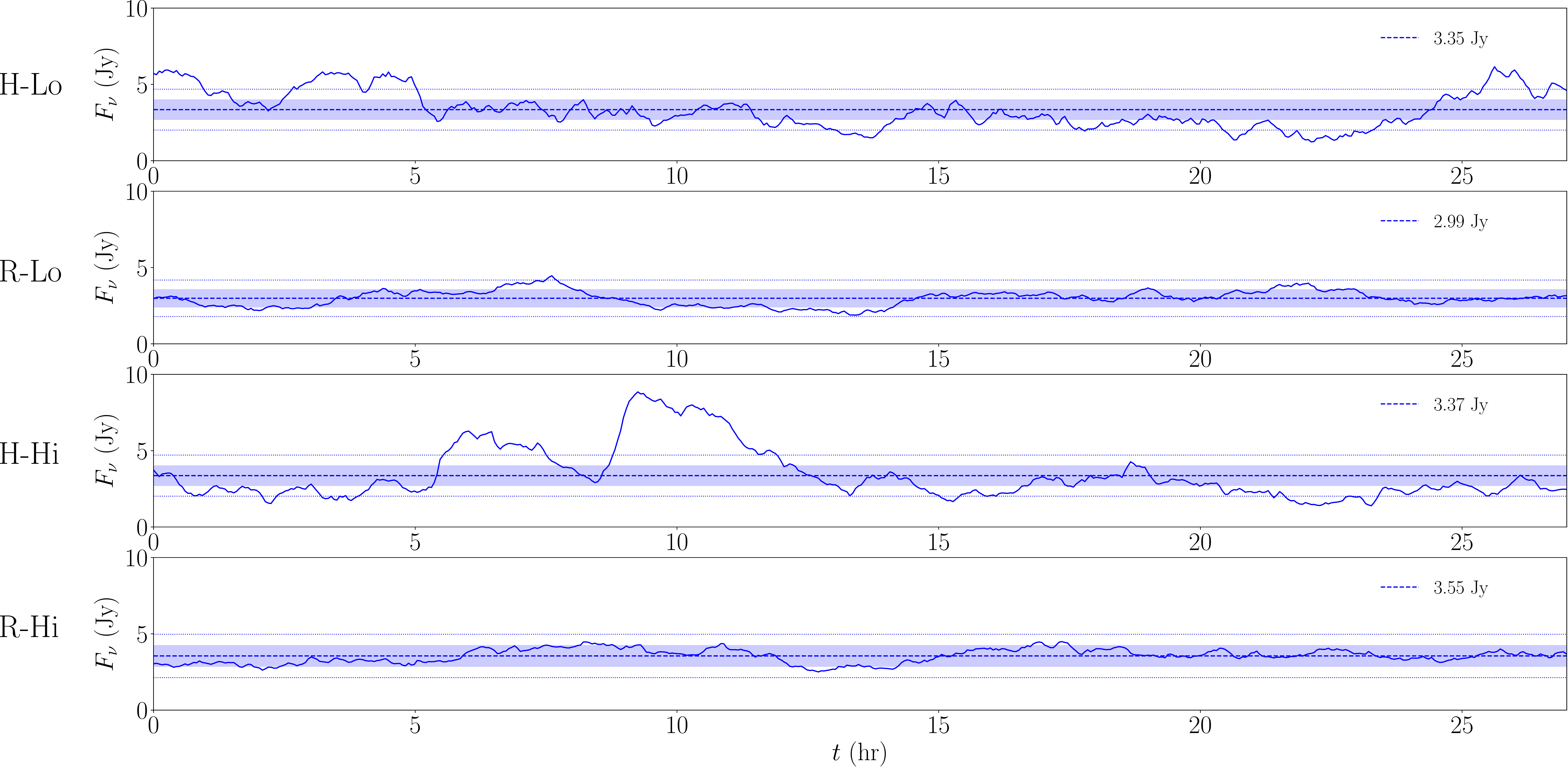}
\caption{
230~GHz lightcurves of the four models at a viewing angle of $60^\circ$ over intervals of $5000 \, t_{\rm g}$ ($\approx27$ hr). The lightcurves were all normalized to be close the observed average flux density of \sgraa\ \citep[$\approx 3.5$ Jy,][]{Bower2015}. The line shows the mean value for each  lightcurve. The shaded band around the mean corresponds to 20 per cent variability; this is roughly the root-mean-square variability level observed in \sgraa\ \citep{Marrone08} at 230~GHz. Dotted lines denote a range of 40 per cent variability around the mean.  All models show variability on $\sim$hr time-scales. The models heated by magnetic reconnection have variability that falls within the observed 20 per cent range, while the models heated by turbulent dissipation (H-Lo and H-Hi) have larger variability amplitudes. Model H-Hi shows two quasi-`flares' around 5~hr and 10~hr that produce excursions above two times the quiescent value. 
}
\label{fig::submmlc}
\end{figure*}

\begin{figure*}
\centering
\includegraphics*[width=0.95\textwidth]{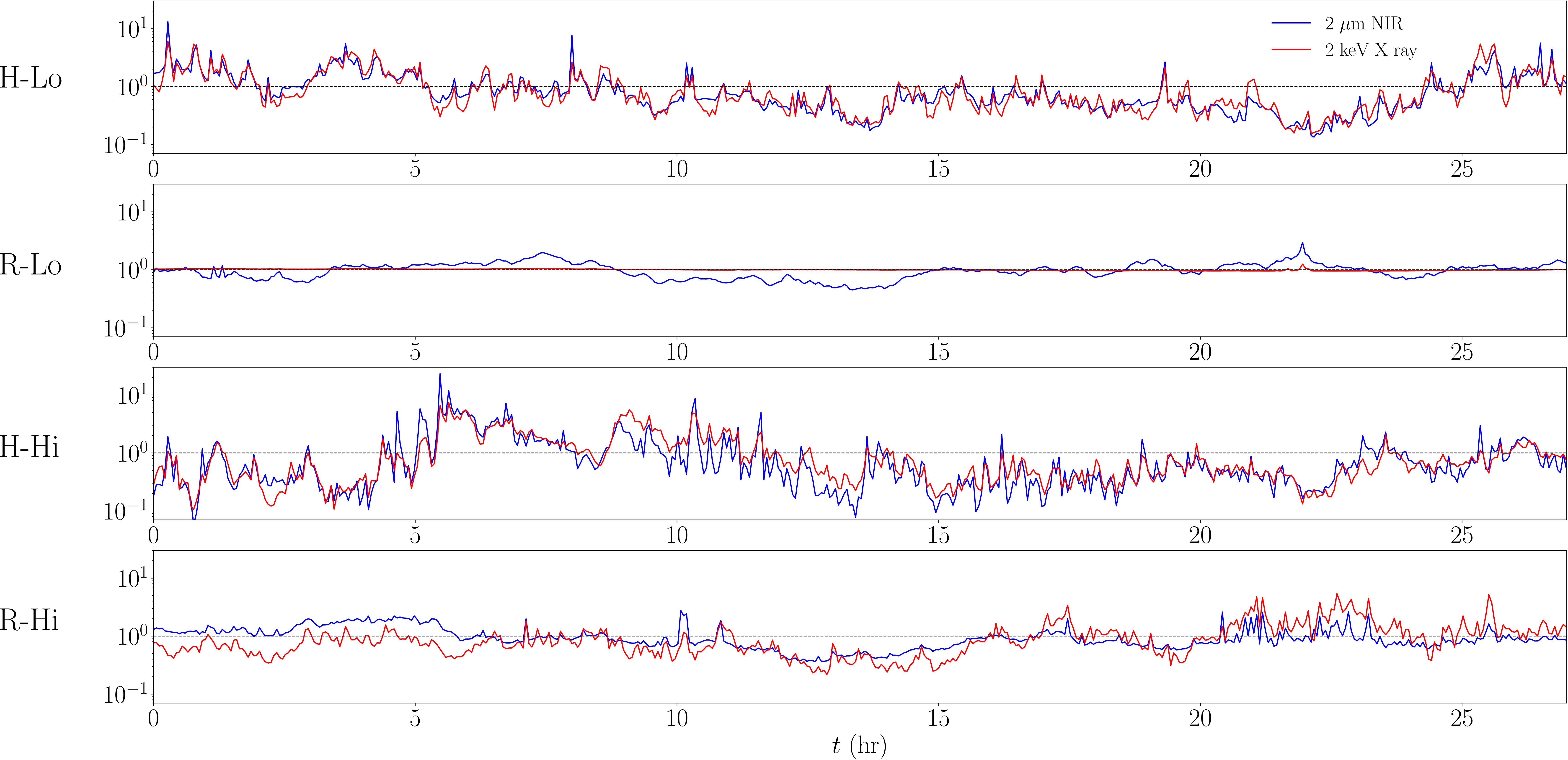}
\caption{
Normalized 2$\mu$m NIR and 2 keV X-ray lightcurves of the four models at a viewing angle of $60^\circ$ over intervals of $5000 \, t_{\rm g}$ ($\approx27$ hr). The curves are normalized to their mean value over the interval. Near-infrared variability arises in thermal synchrotron emission very close to  the black hole, and the variability time-scale is shorter than at 230 GHz. X-ray variability results from inverse Compton scattering of near-infrared photons, and is therefore correlated with the near-infrared variability. In model R-Lo, inverse Compton scattering occurs at low temperatures and does not upscatter  enough photons to 2 keV to outshine the quiescent free-free emission from larger radii. In the other models, all X-ray flaring events have a near-infrared counterpart, but some near-infrared peaks do not get upscattered to 2 keV. We see no strong X-ray flares with amplitudes $>10$ times quiescence. 
}
\label{fig::nirxraylc}
\end{figure*}

Fig.~\ref{fig::submmlc} shows 230~GHz (1.3~mm) light curves for all four models observed at $60^\circ$ inclination, and Fig.~\ref{fig::nirxraylc} shows normalized near-infrared (2~$\mu$m) and X-ray (2 keV) lightcurves over the same time range. The turbulent heating prescription simulations (H-Lo, H-Hi) are more variable than their magnetic reconnection counterparts at all frequencies, since more emission in these models is produced in the high-velocity outflow region away from the equatorial plane, while emission is mostly confined to the disc in the reconnection models (Fig.~\ref{fig::symcompare}). In the extreme case, the spin zero reconnection model R-Lo produces practically no 2 keV X-ray variability, since Compton scattering in the cool disc does not produce enough emission at this frequency to  dominate over the stable quiescent thermal free-free X-ray emission from large radii (See Fig.~\ref{fig::spectra}).

In all cases where they are present, the time variability in the near-infrared and X-ray bands are correlated and more rapidly varying than the millimetre emission. Consistent with past studies \citep{Chan_15b, Ressler17}, we reproduce the result that our X-ray  flaring events all have a near-infrared companion, whereas not all of our near-infrared flares are also seen in X-rays. This matches one qualitative result from observations \citep{Yusef09,Eckart2012}, and is explained in our simulations by X-ray flares being produced by local inverse Compton scattering of near-infrared photons generated by synchrotron emission. However, none of the X-ray flares in our four simulations or in the simulations of \citet{Chan_15b,Ressler17} is anywhere near as bright as those frequently measured  from \sgraa. 

All our models fail to capture other important features of \sgraa's variability. Other than a few large spikes in the near-infrared in model H-Hi, we see no flares more than 10 times brighter than the average, while flares up to $30$ times quiescence are seen in the near-infrared \citep{Dodds11, Witzel2012}. We do not produce any strong X-ray flares with brightness $>10$ times the quiescent flux as are observed on roughly 24 hour time-scales \citep{Neilsen}. As noted in Section~\ref{sec::spectra}, the spectral index of our near-infrared flares is negative in $\nu L_\nu$, while the measured value is positive and seems to be stable over time. \citep{Gillessen2006,Ponti2017}. This positive near-infrared spectral index could be indicative of power-law non-thermal electrons. To produce a positive $\nu L_\nu$ index above the submillimetre peak with only thermal electrons would require that near-infrared emission come from a distinct, stable region, which we do not observe on average in our simulations. Furthermore, measurements of the near-infrared and X-ray spectral indices suggest a cooling break between these bands, indicating that the flaring emission in these bands is synchrotron emission from a power-law non-thermal distribution \citep{Marrone08, Dodds09, Ponti2017}. 

The 230~GHz variability is less pronounced in all cases than the corresponding near-infrared emission, with variability occurring on longer time-scales ($\gtrsim$ 1 hr). The variability amplitude is also less than at shorter wavelengths. The models heated by magnetic reconnection (R-Lo, R-Hi) produce less 230~GHz variability, as their emission is constrained to the less-active disc midplane. The variability in these models falls under the roughly $\sim20$ per cent intraday root-mean-square level observed at 230~GHz \citep{Marrone08, Yusef09, Bower2015}. In contrast, both the models heated by the turbulent heating prescription (H-Lo, H-Hi) show significantly more than the 20 per cent variability observed in \sgraa. Model H-Hi is the most variable at  230~GHz, showing two large excursions with amplitude more than 2 times the average level; these result from sudden activity as material is ejected along the relatively powerful jet. Such dramatic intraday 230~GHz flares have not been observed from \sgraa. 

\subsection{Images}
\label{sec::images}

\begin{figure*}
\centering
\includegraphics*[width=0.99\textwidth]{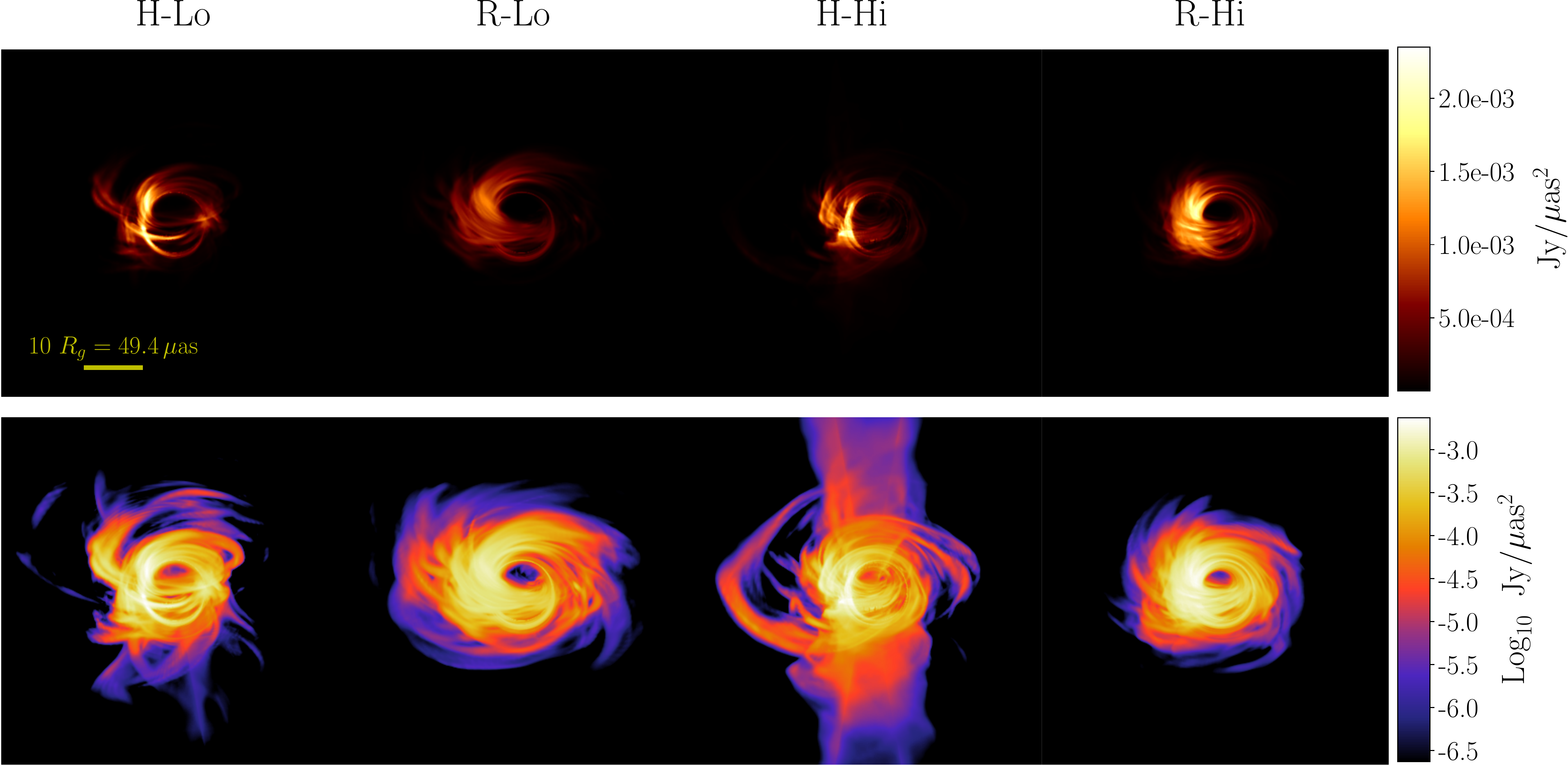}
\caption{
230~GHz (observing frequency of the Event Horizon Telescope) snapshot images at $60^\circ$ inclination. The top row shows images with a linear scale, while the bottom row uses a log scale with a dynamic range of $10{,}000$. The black hole shadow is apparent in all linear scale images. In log scale, the models heated by the turbulent cascade model show emission in the outflow/jet around 100--1000 times fainter than the Doppler-boosted disc emission, while the magnetic reconnection models have all their emission confined to the disc.
}
\label{fig::images}
\end{figure*}

\begin{figure*}
\centering
\includegraphics*[width=0.99\textwidth]{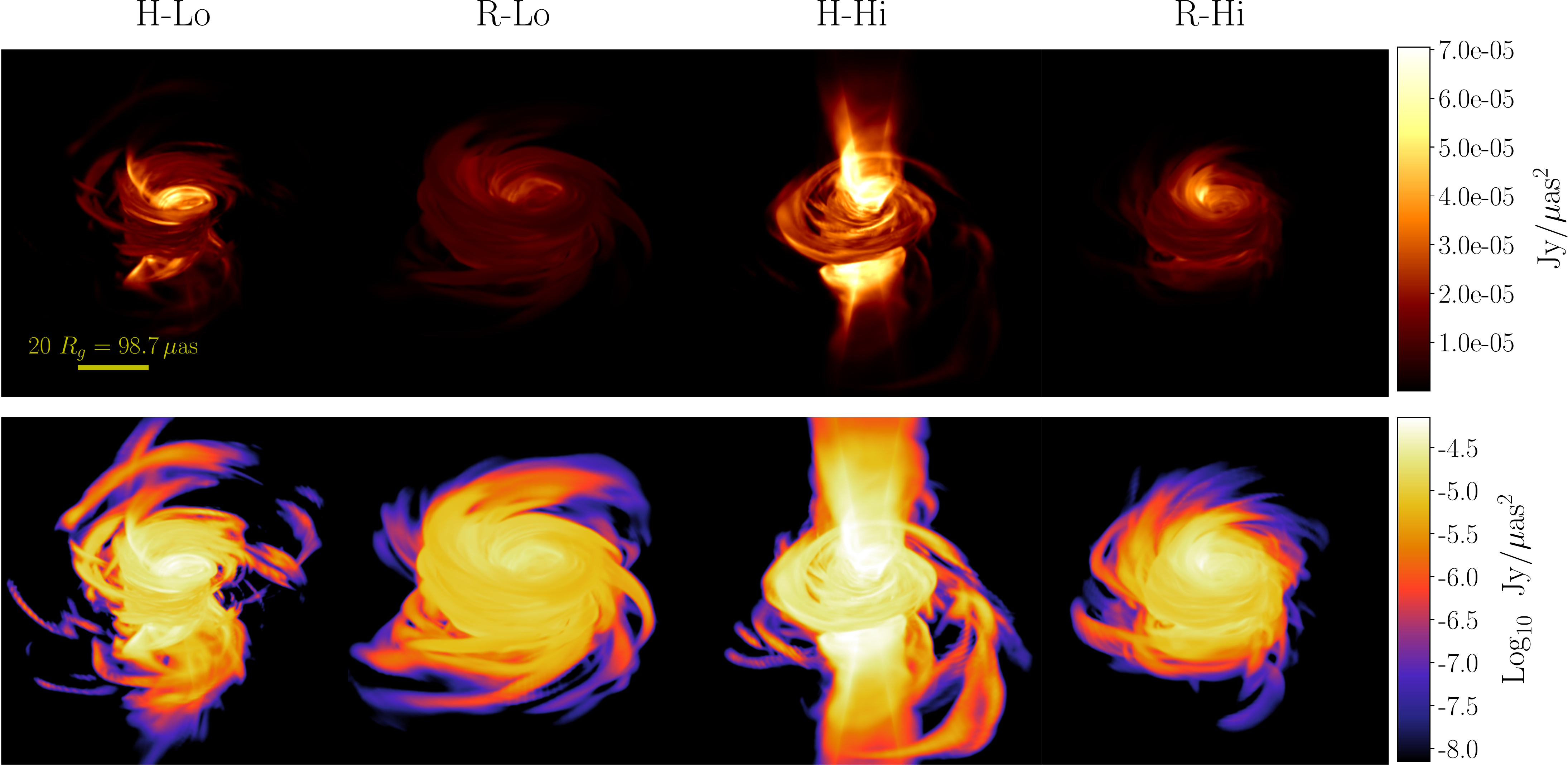}
\caption{
The same snapshots presented in Fig.~\ref{fig::images} at 43~GHz and $60^\circ$ inclination. The top row shows images with a linear scale and the bottom row with a log scale. At this lower frequency, the models heated by the damped turbulent cascade prescription show a pronounced jet, particularly the high spin model H-Hi, which launches a mildly relativistic jet. The magnetic reconnection heated discs produce larger, dimmer images at this wavelength, with emission produced only in the thick disc. 
}
\label{fig::images2}
\end{figure*}

\begin{figure*}
\centering
\includegraphics*[width=0.95\textwidth]{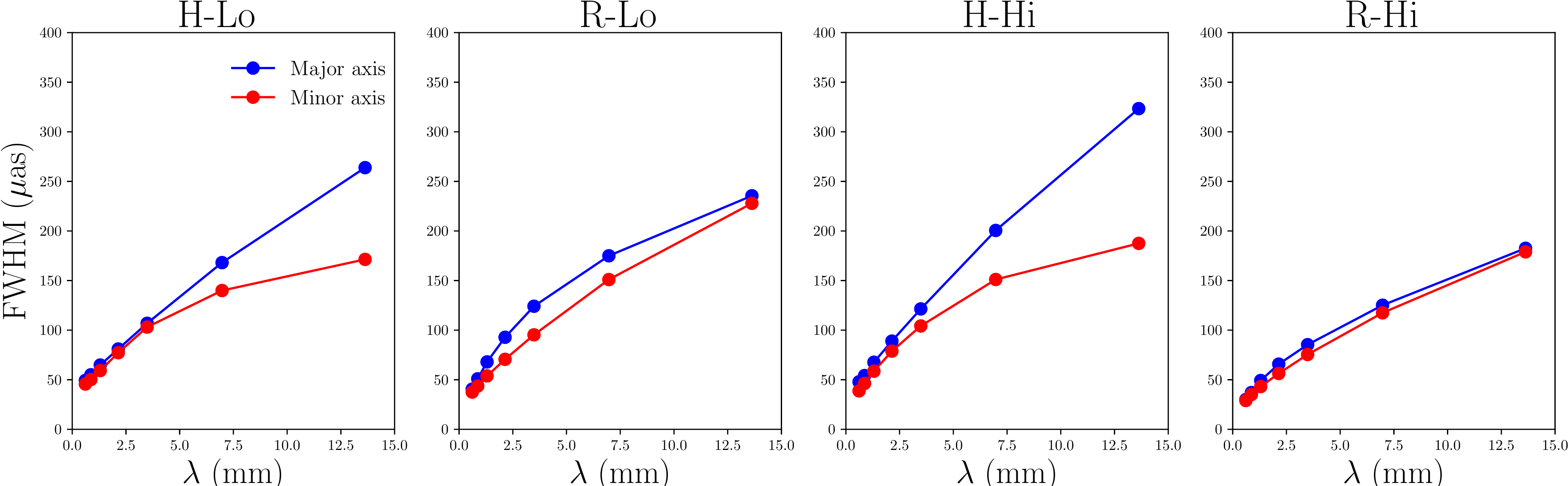}
\caption{
Average image sizes  as a function of wavelength for the four models at $60^\circ$ inclination. Images were averaged over $5000 \, t_{\rm g}$ and the image size in the major and minor axis were calculated by fitting an elliptical Gaussian to the image Fourier transform. The major axis FWHM data are plotted in blue and the minor axis data are plotted in red.  From left to right, sizes are presented for models H-Lo, R-Lo, H-Hi, and R-Hi. The image size grows with wavelength in all cases. In the optically thin regime at wavelengths shorter than 1.3 mm, all the models behave similarly, with similar image sizes growing linearly with wavelength. At longer wavelengths, the models using the turbulent heating prescription show a large anisotropy as the jet/polar outflow begins to dominate the emission, while the models heated by magnetic reconnection remain isotropic.
}
\label{fig::image_sizes}
\end{figure*}

\begin{figure}
\centering
\includegraphics*[width=0.425\textwidth]{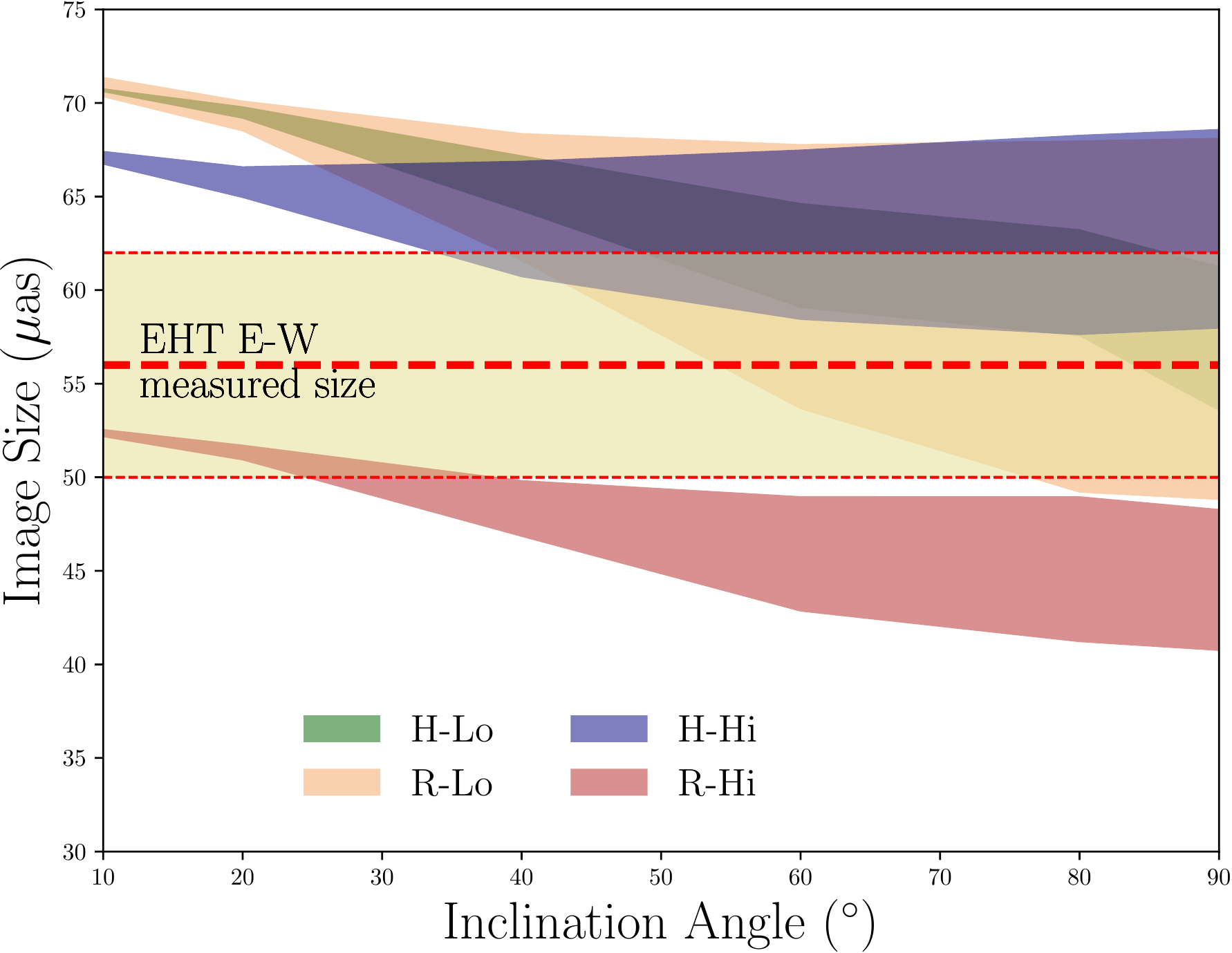}
\caption{
Average image sizes at 230~GHz for the four models as a function of inclination angle. Images were averaged over $5000 \, t_{\rm g}$ and the image size in the major and minor axis were calculated by fitting an elliptical Gaussian to the image Fourier transform. The image sizes are plotted as  bands marking the range of values between the fitted major axis FWHM and minor axis FWHM at each inclination angle. The range of values for the East--West 230~GHz image size measured by the EHT \citep[$56\pm6\,\mu$as][]{Johnson2018} is plotted as a yellow band. In all cases, the image size grows with decreasing inclination angle as Doppler beaming becomes less significant. All four models satisfy the EHT constraint at a given inclination, but model R-Hi only falls within the measurements at nearly face-on inclination. 
}
\label{fig::image_bands}
\end{figure}

Images of our models at $60^\circ$ inclination and 230~GHz, the observing frequency of the EHT, are presented in Fig.~\ref{fig::images}. In all models, the emission at 230~GHz is produced by optically thin synchrotron. The linear scale images are similar among the models. They are brightest on the approaching side of the disc, where emission is relativistically beamed toward the observer. The shadow of the black hole and photon ring are visible in all four models, though slightly more emission emerges from the disc in front of the black hole in models R-Lo and R-Hi. The insensitivity of the appearance of the black hole shadow to the choice of heating prescription in these models is encouraging for the prospects of the EHT to measure the size of the ring and thus test this strong-field prediction of general relativity.

While the linear scale 230~GHz images are similar, the log-scale images reveal significant differences. The models heated by reconnection produce more extended disc emission, and the turbulent cascade heated models produce faint emission in a jet at 230~GHz. Fig.~\ref{fig::images2} shows images of the same snapshots at 43~GHz. At this frequency, the emission is from optically thick synchrotron and the black hole shadow is obscured in all models. The models heated by the turbulent cascade prescription produce most of their 43~GHz emission in the jet or outflow region at large polar angle. The 0.9375 spin model H-Hi has a strong jet collimated around the polar axis, while the polar outflow in the spin zero model is less collimated but still produces an image elongated along the axis perpendicular to the disc. 

To measure the image size and anisotropy at various wavelengths, we define a representative `size' for each image that is motivated by interferometric observations performed by the VLBA and EHT. Specifically, each baseline joining two sites of an interferometric array samples a visibility $\tilde{I}(\bm{u})$, given by the Fourier transform of the image $I(\bm{x})$ \citep{TMSBook}:
\begin{align}
\tilde{I}(\bm{u}) &= \int d^2\bm{x}\, I(\bm{x}) \, \mathrm{e}^{-2\pi \mathrm{i} \bm{u} \bcdot \bm{x}}.
\end{align}
In this expression, $\bm{x}$ is an angular coordinate on the image measured in radians, and $\bm{u}$ is the baseline vector measured in units of the observing wavelength. On a short baseline that only partially resolves the source, we obtain
\begin{align}
\tilde{I}(\bm{u}) &\approx \int d^2\bm{x}\, I(\bm{x}) \left[ 1 - 2\pi \mathrm{i} \bm{u} \bcdot \bm{x} - 2\pi^2  \left( \bm{u} \bcdot \bm{x} \right)^2 \right].
\end{align}
The term linear in $\bm{u}$ gives a visibility phase slope with baseline length that is proportional to the position of the image centroid. Very Long Baseline Interferometry usually lacks absolute phase referencing, so we can define the image centroid to be at the origin and discard the linear term. Short baselines will then see a quadratic fall in the visibility amplitude $\left| \tilde{I}(\bm{u}) \right|$ with increasing baseline length. The quadratic coefficient is proportional to the second moment of the image projected along the baseline vector direction. Thus, we can define the characteristic size along a specified direction in terms of this second moment. In general, the size is anisotropic and will be defined by a quantity analogous to the image moment of inertia tensor. We follow observational conventions and give the image major and minor axis sizes in terms of the equivalent Gaussian major axis full width at half maximum (FWHM), minor axis FWHM, and position angle (measured east of north). For example, 
\begin{align}
\label{eq::fwhm}
\theta_{\rm maj} = \sqrt{-\frac{2 \ln(2)}{\pi^2 I_0}\left. \nabla^2_{\bm{\hat{u}}_{\rm maj}} \tilde{I}(\bm{u})\right\rfloor_{\bm{u}=0}},
\end{align}
where $\theta_{\rm maj}$ is the characteristic FWHM of the major axis, $I_0 \equiv \tilde{I}(\bm{0})$ is the total flux density of the image, and $\nabla^2_{\bm{\hat{u}}_{\rm maj}}$ is the second directional derivative along the direction of the major axis.

For each model, we generate \texttt{grtrans} synchrotron images at 22, 43, 86, 240, 230, 345, and 490~GHz and time average them over the $5000 \, t_{\rm g}$ range considered (Table~\ref{tab::summary}). While we do not reproduce the low-frequency spectrum in any of our models, all the frequencies we examine here are high enough to approximately match the measured spectrum of \sgraa\ (see Fig.~\ref{fig::spectra}), although at 22~GHz the divergence of the model spectra from the data starts to be noticeable. We then determined the image size and orientation according to the definition above (equation~\ref{eq::fwhm}).  

In Fig.~\ref{fig::image_sizes}, we show the resulting FWHMs of the major and minor axes for the four models observed at $60^\circ$ inclination. We see that in the optically thin regime at wavelengths shorter than 1.3~mm, all models produce approximately isotropic images that  grow linearly with wavelength. At longer wavelengths in the optically thick regime, the models heated via the turbulent cascade prescription (H-Lo, H-Hi) show a large anisotropy as the jet or polar outflow begins to dominate the emission. In contrast, the models heated by magnetic reconnection (R-Lo, R-Hi) remain isotropic.

The  transition to jet-dominated emission at low frequencies in the turbulent heating models results from the way the \citet{Howes10} prescription puts most of the dissipated energy into the electrons in the polar regions (Figs.~\ref{fig::symcompare} and \ref{fig::symcompare_curves}). This `disc-jet' structure is consistent with \citet{Ressler17}, who used the same heating prescription (in a more magnetized system). This `disc-jet' has been applied in previous phenomenological models \citep[e.g.][]{Falcke95,Yuan2002} and has been applied to GRMHD simulations by setting the electron temperature in post-processing \citep[e.g.][]{Moscibrodzka_13,Moscibrodzka_14}. This structure is \emph{not} present in our models heated by magnetic reconnection, which does not deposit enough heat in the polar regions to allow the low-density fluid there to make a substantial contribution to the emission. Instead, the 43~GHz emission in these models is confined to the disc and the image is roughly circular when observed at $60^\circ$. Unlike the emergence of the black hole shadow in the optically thin synchrotron emission around 230~GHz, the `disc-jet' structure at low frequencies is not a generic prediction of our GRMHD  models, but is dependent on the choice of heating prescription. 

We can compare the image size predictions from our models with interferometric measurements of \sgraa\ made over the same frequency range. However, these comparisons are complicated by the effects of strong interstellar scattering for the line of sight to \sgraa, with angular broadening from scattering dominating over intrinsic structure at wavelengths longer than a few millimeters. While many authors have inferred the intrinsic size of \sgraa\ at millimetre and centimetre wavelengths by deconvolving these scattering effects, uncertainties in the scattering kernel render these estimates highly uncertain \citep[see, e.g.,][]{Psaltis2015}. The most secure image size estimates are those made with the EHT at 1.3mm, where the scattering effects are minimal, but these suffer from the additional limitation of extremely sparse baseline coverage. While early estimates of the source size found a size of approximately $40\,\mu{\rm as}$ from a direct Gaussian fit to the data, more recent data have found visibility amplitudes on shorter baselines that are widely discrepant from the Gaussian fit \citet{Johnson2015,Lu_2018}. Thus, the appropriate representative image size at 1.3~mm should be estimated by taking the second moment of models that do fit the short- and intermediate-baseline data. For instance, the annulus model from \citet{Johnson2015} gives a size along the East-West direction of $58~\mu{\rm as}$. For comparison, the annulus model from \citet{Doeleman08} gives a characteristic size of $51~\mu{\rm as}$. Following \citet{Johnson2018}, we adopt $50-62\mu{\rm as}$ as a representative range of image size along the East-West direction, as constrained by current EHT data. 

In Fig.~\ref{fig::image_bands}, we show average image sizes fit to the time-averaged images at 230~GHz for our four models as a function of inclination angle. The image sizes are plotted as  bands marking the range of values between the fitted major axis FWHM and minor axis FWHM at each inclination angle. In all our models at 1.3 mm, the image size grows with decreasing inclination angle as Doppler beaming becomes less significant. All four models satisfy the EHT constraint for at least one inclination angle. While  models R-Lo, H-Lo, and H-Hi produce sizes consistent with observations at inclinations higher than $45^\circ$, model R-Hi produces the smallest images and only falls within the measurements at nearly face-on inclination. 

\section{Discussion}
\label{sec::discussion}
\subsection{Comparison to Ressler et al. 2017}
\label{sec::ress}
Prior to the present work, \citet{Ressler17} presented a GRMHD simulation of \sgraa\ with a black hole spin of 0.5, using the \citet{Howes10} turbulent cascade  prescription to heat the electrons. Our simulation method differs from that of \citet{Ressler17} in some notable ways. Their work includes the anisotropic conduction of heat along magnetic field lines, which we ignore, although they report this conduction has little effect on the spectrum and image of \sgraa. \citet{Ressler17} ignore the radiative cooling of electrons and Coulomb coupling of electrons to ions, while we include both. Again, these effects are mostly unimportant for very low accretion rate systems like \sgraa. Note that radiative cooling in particular will become significant at higher accretion rates $\gtrsim 10^{-6}\,\dot{M}_\mathrm{Edd}$, and Coulomb coupling at higher densities and accretion rates.  

Most notably, \citet{Ressler17} uses a fixed adiabatic index $\Gamma_\mathrm{gas}=5/3$ in evolving the total gas, from which the dissipation is identified. In a separate post-processing step they set $\Gamma_\mathrm{e}=4/3$ in evolving the electrons and estimating their temperature. In the trans-relativistic regime of the accretion flow in \sgraa, electrons transition from non-relativistic ($\Theta_\mathrm{e} < 1, \Gamma_\mathrm{e}\approx5/3$) at large radii to relativistic ($\Theta_\mathrm{e} > 1, \Gamma_\mathrm{e}\approx4/3$) at radii close to the black hole and in the outflow. As a result, the effective adiabatic index of the total gas (equation~\ref{eq::gammaeff}) will not be fixed at $5/3$, even if electrons are cooler than ions or have less than 50 per cent of the thermal energy (see Fig.~\ref{fig::symcompare2}). Changes in the effective adiabatic indices in different regions of the simulation will affect the thermodynamics and the amount of dissipation identified in the numerical evolution. For instance, in the simple analytic shock test model presented in \citet{Ressler15}, a gas with an adiabatic index $\Gamma_\mathrm{gas}<5/3$ produces more dissipation and heats electrons to higher temperatures. This difference could be important, especially in the jet region where $\adi$ is well below $5/3$. The different treatment of the species and total gas adiabatic indices presented in \citet{Ressler15, Ressler17} versus \citet{KORAL16} and the present work, and the different effects on the amount of dissipation identified between the treatments, deserves further study, particularly in more magnetized systems (see Section~\ref{sec::mag}).

Despite the various differences in approach described above, the picture from the low and high spin turbulent cascade models (H-Lo, H-Hi) in the present work and the model presented in \citet{Ressler17} (at spin 0.5) is largely consistent. All three models produce similar correlated near-infrared  and X-ray variability from synchrotron self-Compton, and obtain a consistent spectrum at millimetre wavelengths. All these models show more variability at 230~GHz than the approximately 20 per cent observed. In the 230~GHz images, both \citet{Ressler17} and our turbulent cascade models show a pronounced photon ring and  a `disc-jet' structure, where lower frequency emission is dominated by a jet or polar outflow. In this outflow, electrons are heated to high temperatures due to the strong $\beta$-dependence of the \citet{Howes10} turbulent heating prescription. 

\citet{Ressler17} note that using the \citet{Howes10} turbulent heating prescription self-consistently produces the `disc-jet' morphology that had been invoked in previous studies (\citet{Falcke95,Yuan2002,Moscibrodzka_14,Chan_15a}) to explain the low-frequency \sgraa\ spectrum. The main strength of these disc-jet phenomenological models is in reproducing the radio spectrum with an isothermal jet. Earlier works have produced a `disc-jet' structure by setting the electron temperature manually in post-processing. For instance, both \citet{Moscibrodzka_14} and \citet{Chan_15a} identify `jet' and `disc' regions in their single temperature GRMHD simulations based on some criteria and then apply a constant $T_\mathrm{e}$ in the jet and a constant ratio $T_\mathrm{e}/T_\mathrm{i}$ elsewhere. Although a jet is visible in images at frequencies $<230$~GHz in models heated by the \citet{Howes10} turbulent heating prescription, none of the self-consistent thermodynamic models presented in \citet{Ressler17} or the present work reproduce an \emph{isothermal} outflow. 

In this work, we have shown that when we use a different physically motivated heating prescription, namely magnetic reconnection, the disc-jet structure vanishes, at least in the thermal emission. Thus, `disc-jet' morphology is not a guaranteed outcome of simulations of \sgraa\ with self-consistent electron heating. The form of the heating is important in determining the image shape and evolution with  wavelength. Even when the `disc-jet' structure is present, it remains unclear how to provide the jet with the additional heating needed for it to remain isothermal. 

\subsection{Disc Magnetization}
\label{sec::mag}

A key difference between the spectra of our models using the \citet{Howes10} turbulent heating prescription and the spectrum presented by \citet{Ressler17} is that their model produces substantially more near-infrared synchrotron emission, and meets (or even exceeds) measurements of the quiescent near-infrared and X-ray emission. This is most likely because they consider a disc that is substantially more magnetized than ours. The dimensionless magnetic flux $\Phi_{\rm BH} / (\dot{M}c)^{1/2} r_{\rm g} \,\approx \, 40 $ in their model, which is close to the MAD saturation value of $\approx50$ \citep{Tchekhovskoy11}. In contrast, our models all have $\Phi_{\rm BH} / (\dot{M}c)^{1/2} r_{\rm g} \,<\, 10$ (see Table~\ref{tab::summary}). 

As a result, when compared to our models H-Lo and H-Hi, \citet{Ressler17}'s simulation has much lower $\beta_\mathrm{i}$ and a much higher $\sigma_\mathrm{i}$ in the outflow and close to the black hole. Whereas we only see $\sigma_{\rm i}>1$ at the innermost radii in our high spin model H-Hi, \citet{Ressler17} find large $\sigma_{\rm i}>1$ in a substantial part of the outflow closest to the axis (though they exclude this region from their radiative transfer). In their model, $\beta_\mathrm{i}$ averaged over the inner $25 \, r_{\rm g}$ drops below $10^{-1}$ at polar angles $<30^\circ$ and $>150^\circ$, while we only see similar behavior at $5 \, r_{\rm g}$ in the spin 0.9375 model. Consequently, the heating rate $\delta_\mathrm{e}$ in their model is greater at a given radius and polar angle than in our simulations H-Lo and H-Hi. Temperatures in their model reach $\Theta_\mathrm{e}\approx100$, whereas our maximum $\Theta_\mathrm{e}$ is 30 in model H-Hi (see Fig.~\ref{fig::symcompare_curves}). 

This combination of hot electrons and strong magnetic fields in the  inner disc and outflow combine to produce more near-infrared synchrotron in the \citet{Ressler17} simulation, and the median spectrum presented in their work goes though the measured quiescent values from \sgraa. However, just as in our models, their simulation fails to reproduce the measured near-infrared flare spectral slope ($\nu L_\nu\propto \nu^{0.3}$)  or the large observed flare amplitudes in both the near-infrared and X-ray. In fact, the normalized variability in their models is quite similar to  our model H-Lo in the millimetre, near-infrared, and X-ray. In the 230~GHz emission, both our model H-Lo and their model show variability amplitudes on the order of 40 per cent relative to the mean, which is significantly larger than the observed root-mean-square range of of 20 per cent \citep{Marrone08}. In the near-infrared and X-ray, all excursions are contained within one order of magnitude from the mean and no strong flares are generated. 

When the near-infrared quiescent emission in \citet{Ressler17}'s simulation is inverse Compton upscattered to X-ray frequencies, it results in more quiescent X-ray emission than we see in any of our models, at the upper limit of the quiescent range of 10 -- 100 per cent of the \citet{Baganoff03} value. This is despite the fact that \citet{Ressler17} do not include bremsstrahlung emission. It seems likely that if this were included, their model would overpredict the total measured \sgraa\ quiescent X-ray emission. As the turbulent heating prescription puts nearly 100 per cent of the energy into electrons in the jet and close to the black hole, at higher disc magnetizations the gas adiabatic index $\adi$ will become closer to $4/3$ than $5/3$ in a substantial part of the accretion flow. In this regime, the self-consistent treatment of the adiabatic index $\adi$ used in \texttt{KORAL} could become important and lead to different results for the jet luminosity and spectrum from those reported in \citet{Ressler17}. 

\subsection{The need for a non-thermal population}
\label{sec::nth}

Our four models all produce spectra that match observations of \sgraa\ at frequencies near the synchrotron peak around $10^{11}$--$10^{12}$~Hz (Fig.~\ref{fig::spectra}). In addition, they all produce 230~GHz images consistent with the size measured by the EHT over some range of inclination angle (Fig.~\ref{fig::image_bands}). However, none of our models reproduce the characteristic large-amplitude X-ray flares observed $\sim$daily from \sgraa, they all underpredict the quiescent near-infrared emission. In addition, they do not show bright infrared flares with hard spectra, and they fail to reproduce the low-frequency radio spectral slope. 

While an isothermal jet can fit the low-frequency radio data \citep{Yuan2002, Moscibrodzka_14, Chan_15a}, a high-energy non-thermal electron population is another potential solution \citep{Ozel2000, Yuan2003}. Recently, \citet{Davelaar_17} have applied a hybrid non-thermal-thermal $\kappa$ distribution function in the jet in postprocessing \sgraa\ GRMHD simulations. They show that non-thermal electrons in the jet can match both the relatively flat low frequency spectrum and the measured near-infrared spectral index. This effectively recovers the `disc-jet' model, but lights up the jet with non-thermal particles instead of hot, thermal electrons, similar to the original disc-jet model of \citet{Falcke95}.  

No thermal-only model has successfully reproduced the observed infrared variability or X-ray flares from \sgraa. \citet{Chan_15b}, \citet{Ressler17}, and the present work all  reproduce the observed qualitative behavior whereby spikes in the X-ray always correspond to a near-infrared event, but the converse is not always true. This behavior is a natural result of synchrotron self-Compton, whereby the X-ray flares are generated by inverse Compton upscattering from near-infrared synchrotron photons. However, neither this work nor previous works have successfully reproduced the large flare amplitudes observed in the X-ray and near-infrared, nor the positive $\nu L_\nu$ power-law slope measured in the near-infrared \citep{Genzel03,Gillessen2006,Hornstein2007}. The positive spectral index is a particularly important clue pointing toward non-thermal electrons, as no thermal synchrotron model that peaks in the submillimetre can produce a positive spectral index in the near-infrared.  Furthermore, \citet{Marrone08, Dodds09, Ponti2017} report a spectral index difference of $\approx 0.5$ between the X-ray and near-infrared, suggestive of a synchrotron cooling break between the near-infrared and X-ray.  

The large amplitudes of the observed near-infrared and X-ray flares again point to non-thermal electrons. \citet{Ball_16} demonstrated that inserting localized patches of non-thermal electrons in post-processing can produce strong X-ray flares of greater than 10 times the quiescent value. \citet{Li2017} used an analytic MHD model to show that magnetic reconnection of flux ropes powering the acceleration of non-thermal electrons can reproduce the main features of near-infrared and X-ray flares from non-thermal synchrotron radiation. Cooling non-thermal electrons in a strong $B$ field also provides an alternative explanation to synchrotron self-Compton for the observed correlations between X-ray and near-infrared flares and the shorter lifetimes of the X-ray flares \citep{Kusunose11}. 

To properly explore the signatures of non-thermal emission one should include non-thermal particle acceleration and self-consistent evolution in the GRMHD simulation. \citet{Chael17} have developed a code that does precisely this. In a forthcoming work, we will explore the effects of physically-motivated sub-grid acceleration models (such as particle acceleration from reconnection, observed in trans-relativistic simulations by \citet{Werner2018, Ball2018}. We  will use physically motivated models to accelerate a fraction of electrons into power-law distributions in \sgraa simulations, and explore their evolution and effects on the radio spectral slope, near-infrared and  X-ray variability, and near-infrared spectral index.  

\section{Summary}
\label{sec::summary}

We have performed four two-temperature GRRMHD simulations of \sgraa\ using different combinations of black hole spin and electron heating prescriptions. Our heating prescriptions are motivated by different models for the plasma microphysics around \sgraa. We have introduced a new heating prescription (equation~\ref{eq::Rowan}) based on PIC simulations of anti-parallel magnetic reconnection presented in \citet{Rowan17}. We compare this model with the Landau-damped turbulent cascade prescription of \citet{Howes10}. These two heating prescriptions have starkly different qualitative features. The turbulent heating prescription is a sharp function of the ratio of the gas pressure to magnetic pressure $\beta_\mathrm{i}$, putting almost all the dissipated energy into electrons at low $\beta_\mathrm{i}$ and almost  all energy into ions at high $\beta_\mathrm{i}$ (Fig.~\ref{fig::howesmodel}). In contrast, the model fit to particle-in-cell simulations of reconnection varies less rapidly with $\beta_\mathrm{i}$ and never puts more than half of the heat into electrons (Fig.~\ref{fig::rowanmodel}).

This difference in the heating prescriptions has major effects on the properties of the accretion flow, as well as on the resulting simulated spectra and images. Under the turbulent cascade heating prescription, even though our simulations have a relatively weak magnetic field, electrons are heated to  very high temperatures in the funnel and are cooler in the disc. In contrast, the reconnection heating prescription heats electrons by nearly the same fraction everywhere (Fig.~\ref{fig::symcompare_curves}). Energy is mostly radiated from the disc in our two  simulations using the reconnection heating prescription, whereas with turbulent heating a significant amount of the radiation comes from the jet and outflow. This is particularly true in the high spin model H-Hi, which launches a mildly relativistic (Lorentz factor $\approx$2) jet (Fig.~\ref{fig::symcompare2}). 

Once normalized to the 230~GHz flux density observed for \sgraa, the spectra of all the four models match observations and are similar over the range $10^{11}$--$10^{12}$ Hz (Fig.~\ref{fig::spectra}). However, none of our thermal models can reproduce  the low frequency radio spectrum nor the near-infrared flux density and spectral index. While the variability from synchrotron self-Compton produces a correlation between the near-infrared and X-ray that is qualitatively similar to the observed behavior, we do not see any large near-infrared or X-ray flares (Fig.~\ref{fig::nirxraylc}). Because more of their emission comes from the outflow and jet, the models heated by the  turbulent cascade prescription are highly variable, and exceed the 20 per cent level of root-mean-square variability measured for \sgraa.  The models heated by reconnection, on the other hand, all lie within the 20 per cent variability bands at 230~GHz (Fig.~\ref{fig::submmlc}).

We found that all four models produce 230~GHz images with distinct shadows and photon rings. All models produce average 230~GHz images that are consistent with the size measured by the EHT over some range of inclination. Consistent with past studies, the turbulent heating prescription simulations produce images that are dominated by an outflow or jet at frequencies lower than 230~GHz. In contrast, neither simulation using the magnetic reconnection heating prescription produces a jet in the image at lower frequencies (Figs.~\ref{fig::images2} and~\ref{fig::image_sizes}). Thus we conclude that while the transition of the synchrotron emission from optically thick to optically thin and the emergence of the black hole shadow around 230~GHz is a universal feature in all our models, a `disc-jet' structure is not. It is sensitive to the choice of thermal electron heating prescription. 

Our work explores only weakly magnetized discs, and further simulations must be performed to compare different heating mechanisms in discs at or near the MAD limit. However, while more magnetized simulations may produce higher near-infrared and X-ray quiescent flux, simply taking our thermal two-temperature simulation to greater magnetizations is unlikely to produce either the correct radio or near-infrared spectral indices or strong X-ray flares. Recent work in adding non-thermal electron distributions to GRMHD simulations in postprocessing \citep{Ball_16, Davelaar_17} has supported earlier analytic work \citep{Ozel2000, Yuan2003, Kusunose11} indicating that high-energy non-thermal populations are necessary to solve these remaining problems in modelling \sgraa's spectrum and variability. Electron acceleration to non-thermal energies is observed in particle-in-cell simulations of trans-relativistic reconnection \citep{Werner2018, Ball2018}. In a future work, we will couple the self-consistent non-thermal electron evolution method developed in \citet{Chael17} with physical models of relativistic, non-thermal electron acceleration. 

\section*{Acknowledgements}
We thank the referee, Sean Ressler, for his helpful comments and exceptionally prompt review. 
We thank Jason Dexter for his help in using and modifying \texttt{grtrans}.
This work was supported in part by NASA via the TCAN award grant NNX14AB47G. 
AC was supported in part by NSF grant AST-1440254. 
RN was supported in part by NSF grant AST-1312651. 
MJ was supported by NSF grant AST-1716536.
The authors acknowledge
computational support from NSF via XSEDE resources (grant TG-AST080026N). 
This work was conducted at the Black Hole Initiative at Harvard University, supported by a grant from the John Templeton Foundation.

\appendix 

\section{Thermodynamic Relations}
\label{appendix::eqs}
Here we present the thermodynamic relations used in \texttt{KORAL} which relate the species temperatures, pressures, and energy densities. The following 
equations are applicable for both ions and electrons, using the appropriate mass $m_{i,e}$. From charge neutrality, the electron and ion number densities 
are equal $n_\mathrm{e} = n_\mathrm{i} = n = \rho/m_\mathrm{i}$.   

The ions and electrons are considered to each be in relativistic, nondegenerate thermal Maxwell-J\"{u}ttner distributions. At a given temperature $T$ and 
number density $n$, the pressure and energy density for a given species are \citep{Chandra39}
\begin{align}
\label{eq::perfectgas}
p &= n k_{\rm B} T, \\
u &= \frac{p}{\Gamma(\Theta)-1} \\
\Gamma(\Theta) - 1 &= \Theta\left(\frac{3K_3\left(1/\Theta\right) + K_1\left(1/\Theta\right)}{4K_2\left(1/\Theta\right)}-1\right)^{-1}, 
\end{align}
where $\Gamma$ is the the adiabatic index
$\Theta = k_{\rm B} T/m c^2$ is the dimensionless temperature, and $K_n(x)$ is the modified 
Bessel function of order $n$. The classical entropy per particle for the Maxwell-J\"{u}ttner distribution is
\begin{equation}
 s/k_{\rm B} =  \frac{K_1\left(1/\Theta\right)}{\Theta K_2\left(1/\Theta\right)} + \ln\left[\frac{\Theta K_2\left(1/\Theta\right)}{n}\right] + C,
\end{equation}
where $C$ is an integration constant. 

Because the exact expressions above involve expensive Bessel functions and are difficult to invert, 
we use approximate forms when evolving equations~\eqref{eq::ent_ev} and~\eqref{eq::ent_ev2} in \texttt{KORAL}. Our self-consistent approximation is based on a fitting function to the specific
heat at constant volume, which we can integrate to find expressions
for the internal energy and entropy per particle (see Appendix A of \citealt{KORAL16}). Our approximate equation 
for the adiabatic index and internal energy is:
\begin{equation}
 \label{eq::adiabspecies}
 \frac{u(\Theta)}{p(\Theta)} = \frac{1}{\Gamma(\Theta) - 1} \approx 3-\frac{3}{5\Theta}\ln\left[1+\frac{5\Theta}{2}\right],
\end{equation}
The entropy per particle is 
\begin{equation}
\label{eq::entpp}
s/k_{\rm B} \approx \ln\left[\frac{\Theta^{3/2}(\Theta+2/5)^{3/2}}{n}\right] + C,
\end{equation}
where we set the arbitrary integration constant $C=0$. 

Equation~\eqref{eq::entpp} is easy to invert to get the species dimensionless temperature $\Theta$:
\begin{equation}
 \label{eq::thetafroment}
 \Theta \approx \frac{1}{5}\left(\sqrt{1+25\left[n\exp{\frac{s}{k_{\rm B}}}\right]^{2/3}}-1\right).
\end{equation}
Equation~\eqref{eq::adiabspecies} cannot be analytically inverted to solve for $\Theta$ from $u$. 
Since this inversion only needed a few times per timestep, we follow \citet{Chael17} and 
use a Newton-Raphson solver to invert equation~\eqref{eq::adiabspecies} when necessary.  
This differs from the treatment in \citet{KORAL16} which used a simpler but inconsistent
fitting function for $u(\Theta)$ which could be directly inverted.

Because the separate species pressures and energy densities must add to the total gas pressure and energy density ($p_\mathrm{i} + p_\mathrm{e} = p$, 
$u_\mathrm{i} + u_\mathrm{e} = u$), the gas temperature and adiabatic index are (for pure ionized hydrogen)
\begin{align}
\label{eq::gammaeff} 
T_\mathrm{gas} &= \frac{1}{2}\left(T_\mathrm{i} + T_\mathrm{e}\right) \\
\adi -1 &= \frac{\left(\Gamma_\mathrm{i} -1\right)\left(\Gamma_\mathrm{e} -1\right)\left(T_\mathrm{i}/T_\mathrm{e} + 1\right)}{\left(T_\mathrm{i}/T_\mathrm{e}\right)\left(\Gamma_\mathrm{e}-1\right) + \left(\Gamma_\mathrm{i} -1\right)}.
\end{align}

\section{Coordinate system and initial conditions}
\label{appendix::grid}
\begin{figure}
\includegraphics*[width=0.45\textwidth]{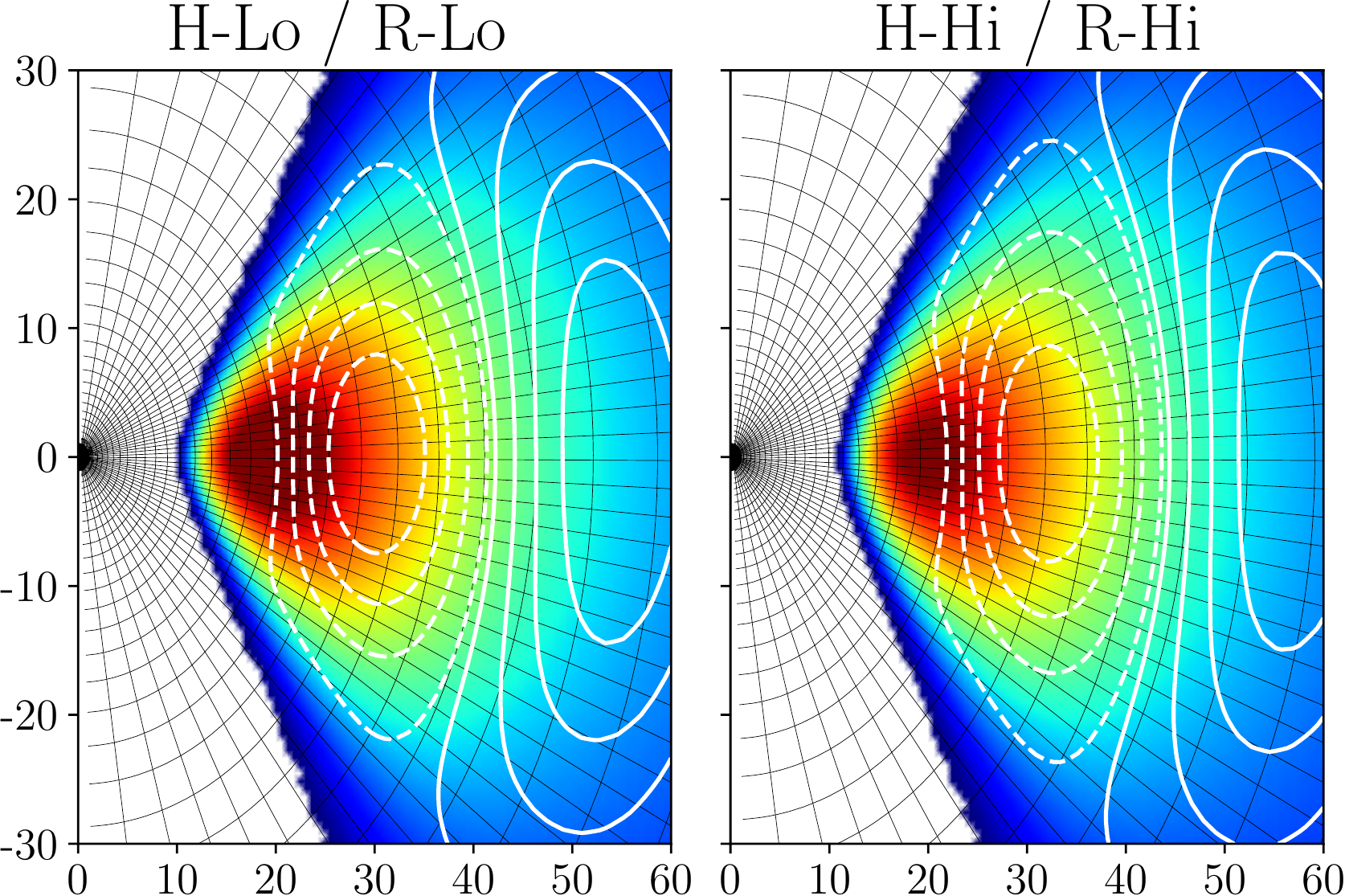}
\caption{
Coordinate grid and initial torii for the two spin zero models R-Lo and H-Lo (left), and for the two spin 0.9375 models R-Hi and H-Hi (right). White contours indicate magnetic field lines.  
}
\label{fig::init}
\end{figure}
Our grid is defined by a mapping that takes our code coordinates $x_1, x_2, x_3$ is to standard Kerr-Schild coordinates \citep{Gammie03} ($r$, $\theta$, $\phi$) in the Kerr metric. We choose a coordinate mapping that is exponential in $r$ and concentrates grid cells near the equator. We also choose a functional form that naturally `cylindrifies' our grid cells closer to the poles, expanding them laterally at small radii so that the  coordinates in the inner region are more cylindrical than spherical. This greatly speeds up the simulation by limiting the time step constraint imposed by the Courant condition \citep{Tchekhovskoy11}.

Our grid is defined by the equations: 
\begin{align}
 \label{eq::mks3tomks}
 r &= \mathrm{e}^{x_1} + r_0 \nonumber\\
 \theta &= \frac{\pi}{2}\left\{1+\tan\left[\pi h_0 \left(\left(1-2x_2\right)\left(\frac{2^p(y_2-y_1)}{\left(\mathrm{e}^{x_1}+r_0\right)^p}+y_1\right) + \right.\right.\right. \\ 
      &\vphantom{=} \:\:\:\:\:\:\:\:\:\:\:\:\:\:\:\:\:\:\:\:\:\:\:\:\:\:\:\:\:\:\:\:\:\:\:\:\:\:\: \left.\left.\left.\left(x_2 -\frac{1}{2} \right)\right)\right]\cot\left[\frac{\pi h_0}{2}\right]\right\} \nonumber\\
 \phi &= x_3
\end{align}
The parameter $r_0<0$ changes the grid spacing near the origin, with a smaller $|r_0|$ placing more cells near $r_\text{min}$. Increasing parameter $h>0$ concentrates cells toward the equatorial plane. Making $y_1>0$ larger (at fixed $h$) increases the minimum polar angle at large $r$, and increasing $y_2>0$ increases the minimum polar angle at small $r$. Adjusting $p>0$ changes how quickly the minimum polar angle at a given radius changes between the value at $r_min$ and the value at $r_max$. For all our simulations, we fixed $h_0 = 0.7$, $y_2 = 0.02$, $y_1 = 0.002$, $p = 1.3$. For the spin $a=0$ simulations, we chose $r_0=-2$, while for $a=0.9375$, $r_0 = -1.35$.

The initial gas torus was set using the \citet{Penna13} model, which
defines a torus that has an angular momentum profile that is proportional to Keplerian  over a certain radial range $[r_\text{k,min},\,r_\text{k,max}]$. The equatorial plane angular momentum is 
constant outside these limits. The adiabatic index is fixed at $\Gamma_{\rm gas}=5/3$ in setting up the initial torus. The spin zero models R-Lo, H-Lo were initialized with an inner edge at $10 \, r_{\rm g}$, a transition to a angular momentum profile with values $\xi=0.708$ times the Keplerian value in the range $[42 r_{\rm g}, \; 1000 r_{\rm g}]$. The spin 0.9375 torus was nearly identical, except that the strong dependence of the \citet{Penna13} model on spin means that setting the inner edge at $10 r_{\rm g}$ produces a torus that nearly fills the entire grid. To avoid this, we set the inner edge at $11 \, r_{\rm g}$, keeping all other values fixed. 

The initial magnetic field is set up in the torus with alternating dipolar field loops, with the field
strength normalized such that the maximum value of $\beta_\mathrm{i}$ in the midplane is $10^{-2}$. In all models, the initial energy in electrons was taken to be 1 per cent of the total gas energy, with the remainder in ions. The initial torus is surrounded by a static atmosphere with an $r^2$ profile and negligible mass and energy density. The initial radiation energy density is negligible everywhere. 
Our initial torii and simulation grids are displayed in Fig.~\ref{fig::init}.

\section{Observational Data}
\label{appendix::data}
The data points presented in the spectra in Figs.~\ref{fig::spectra} and \ref{fig::spectra_inc} are mostly the same as plotted in the spectra in \citet{Ressler17} with some additions. 

We take radio and millimetre points from \citet{Falcke98}  over the range 1.46--235.6~GHz, from \citet{An2005} over the range 0.33--42.9~GHz, from \citet{Bower2015} in the range 1.6--352.6~GHz and from \citet{Liu2016a} and \cite{Liu2016b} in the interval 93--709~GHz and at 492~GHz, respectively.
230~GHz measurements of the total flux density using the Event Horizon Telescope (EHT) were taken from \citet{Doeleman08} and \citet{Johnson2015}.

Infrared upper limits were taken from \citet{Cotera99} in the range 8.7--24.5~$\mu$m, from \citet{Genzel99} at 2.2~$\mu$m and \citet{Schodel07} at 8.6~$\mu$m. \citet{Genzel03} provide infrared flux density measurements for both the quiescent state and flares at 1.76, 2.16, and 3.76~$\mu$m. \citet{Schodel11} provides quiescent state measurements at 2.1, 3.8, and 4.8~$\mu$m, and \citet{Witzel2012} reports a quiescent value at 2.2~$\mu$m. 

We include a range of X-ray flare luminosities over the range 2-10 keV reported in \citet{Neilsen}, and our measure of the quiescent X-ray luminosity is taken from \citet{Baganoff03}. As \citet{Neilsen} notes that only about 10 per cent of the X-ray quiescent luminosity is produced in the inner accretion flow, we denote the range between 10 per cent and 100 per cent of the \citet{Baganoff03} measurement as the lower shaded band in Figs.~\ref{fig::spectra}~\ref{fig::spectra_inc}. 

Simple estimates of the root-mean-square (RMS) variability in the 230~GHz light curve were plotted as 20 and 40 per cent bands in Fig.~\ref{fig::submmlc}. \citet{Marrone08, Yusef09, Bower2015} all report a value of roughly 20 per cent RMS variability relative to the mean. Finally, the 230 GHz image size estimate in the E-W direction comes from Event Horizon Telescope data reported in \citet{Doeleman08} and \citet{Johnson2015}.

\bibliography{ElecEv}

\end{document}